\documentclass[acmsmall]{acmart}

\AtBeginDocument{%
  \providecommand\BibTeX{{%
    \normalfont B\kern-0.5em{\scshape i\kern-0.25em b}\kern-0.8em\TeX}}}

\setcopyright{rightsretained}
\usepackage[inline]{enumitem}     
\newcommand*\circled[1]{\tikz[baseline=(char.base)]{
            \node[shape=circle,fill,inner sep=0.5pt] (char) {\textcolor{white}{#1}};}}    

\usepackage{graphicx}
\usepackage{subcaption} 

\usepackage{textcomp}
\usepackage{multirow}  
\usepackage{xcolor}
\def\BibTeX{{\rm B\kern-.05em{\sc i\kern-.025em b}\kern-.08em
    T\kern-.1667em\lower.7ex\hbox{E}\kern-.125emX}}
    
\usepackage{hyperref} 
\usepackage{listings}

\usepackage{changepage}

\usepackage{ragged2e}
\usepackage{seqsplit}
\pdfmapfile{=rtxr.map} 

\usepackage[font=footnotesize,skip=1pt,labelfont=bf,tableposition=top]{caption}
\DeclareCaptionFont{white}{\color{white}}
\DeclareCaptionFormat{listing}{%
  \parbox{\textwidth}{\colorbox{gray}{\parbox{\textwidth}{#1#2#3}}\vskip-4pt}}
\hypersetup{colorlinks,linkcolor={blue},citecolor={blue},urlcolor={blue}}

\usepackage{algpseudocode}
\makeatletter
\def\BState{\State\hskip-\ALG@thistlm}
\makeatother
\usepackage{balance}
\usepackage[skins,listings,breakable]{tcolorbox}

\makeatletter
\let\old@lstKV@SwitchCases\lstKV@SwitchCases
\def\lstKV@SwitchCases#1#2#3{}
\makeatother

\usepackage{pdflscape}
\usepackage{rotating}
\usepackage{booktabs}
\usepackage{multirow}

\usepackage{lstlinebgrd}
\usepackage[utf8x]{inputenc}
\usepackage{ucs}
\usepackage{listings}
\usepackage{amsfonts}
\usepackage{tcolorbox}
\newtcbox{\highlight}[0]{boxsep=0pt,left=0pt,top=0pt,bottom=0pt,right=0pt,boxrule=0pt,arc=0pt,auto outer arc,colback=green,width=9cm}

\makeatletter
\let\lstKV@SwitchCases\old@lstKV@SwitchCases
\lst@Key{numbers}{none}{%
    \def\lst@PlaceNumber{\lst@linebgrd}%
    \lstKV@SwitchCases{#1}%
    {none:\\%
     left:\def\lst@PlaceNumber{\llap{\normalfont
                \lst@numberstyle{\thelstnumber}\kern\lst@numbersep}\lst@linebgrd}\\%
     right:\def\lst@PlaceNumber{\rlap{\normalfont
                \kern\linewidth \kern\lst@numbersep
                \lst@numberstyle{\thelstnumber}}\lst@linebgrd}%
    }{\PackageError{Listings}{Numbers #1 unknown}\@ehc}}
\makeatother

\lstdefinestyle{base}{
  language=Java,
  basicstyle=\ttfamily\scriptsize,
  frame=lines,
  keywordstyle=\color{blue}\textbf,
  commentstyle=\color[rgb]{0.0,0.4,0.0}\scriptsize,
  extendedchars=true,         
  breaklines=true   
  showspaces=false,
  showstringspaces=false, 
   numbers=left,
    stepnumber=1,   
        tabsize=1,
    breaklines=true,      
    xleftmargin={0.5cm},
  moredelim=**[is][\color{green}]{!!}{!!},
  moredelim=**[is][\color{orange}]{^}{^},
  moredelim=**[is][\color{red}]{@}{@},
   breakindent=0pt,                % Space before the text of a break
     }

\DeclareCaptionLabelFormat{andtable}{#1~#2  \&  \tablename~\thetable} 

\usepackage{diagbox}

\usepackage[linesnumbered,ruled,vlined]{algorithm2e}

\SetCommentSty{mycommfont}

\SetKwInput{KwInput}{Input}                % Set the Input
\SetKwInput{KwOutput}{Output}              % set the Output

\SetAlgorithmName{Algo.}{}{}
\usepackage{soul}

\colorlet{soulred}{blue!10}
\DeclareRobustCommand{\hlcyan}[1]{{\sethlcolor{soulred}\hl{#1}}}

\usepackage{array}
\newcolumntype{P}[1]{>{\centering\arraybackslash}p{#1}}
\usepackage{colortbl}

\makeatletter
\newcommand{\removelatexerror}{\let\@latex@error\@gobble}
\makeatother

\usepackage{float}
\setlist[itemize]{leftmargin=4mm}

\usepackage{afterpage}

\usepackage{etoolbox}
\makeatletter
\patchcmd{\maketitle}{\@copyrightspace}{}{}{}
\makeatother

\begin{document}

%%
%% The "title" command has an optional parameter,
%% allowing the author to define a "short title" to be used in page headers.
\title{Borrowing from Similar Code: A Deep Learning NLP-Based Approach for Log Statement Automation}

\author{Sina Gholamian}
\email{sgholamian@uwaterloo.ca}
\orcid{0000-0002-6198-0706}
\author{Paul A. S. Ward}
\email{pasward@uwaterloo.ca}
\affiliation{%
  \institution{University of Waterloo}
  \city{Waterloo}
  \country{Canada}
  \postcode{N2L 3G1}
}

%%
%% By default, the full list of authors will be used in the page
%% headers. Often, this list is too long, and will overlap
%% other information printed in the page headers. This command allows
%% the author to define a more concise list
%% of authors' names for this purpose.
\renewcommand{\shortauthors}{Gholamian and Ward}

%%
%% The abstract is a short summary of the work to be presented in the
%% article.
\begin{abstract}
\noindent\makebox[\linewidth]{\rule{\textwidth}{0.3pt}}
Software developers embed logging statements inside the source code as an imperative duty in modern software development as log files are necessary for tracking down runtime system issues and troubleshooting system management tasks.
Prior research has emphasized the importance of logging statements in the operation and debugging of software systems. 
However, the current logging process is mostly manual, and thus, proper placement and content of logging statements remain as challenges. To overcome these challenges, methods that aim to automate log placement and predict its content, \textit{i.e.}, \textit{`where and what to log'}, are of high interest. 
Thus, we focus on predicting the \textit{location} (\textit{i.e.}, where) and \textit{description} (\textit{i.e.}, what) for log statements by utilizing source code clones and natural language processing
(NLP), as these approaches provide additional context and advantage
for log prediction. 
Specifically, we guide our research with three research questions (RQs): \textbf{(RQ1)} how similar code snippets, \textit{i.e.}, code clones, can be leveraged for log statements prediction? \textbf{(RQ2)} how the approach can be extended to automate log statements' descriptions? and \textbf{(RQ3)} how effective the proposed methods are for log location and description prediction? 
To pursue our RQs, we perform an experimental study on seven open-source Java projects. 
We introduce an updated and improved log-aware code-clone detection method to predict the \textit{location} of logging statements (RQ1). 
Then, we incorporate natural language processing (NLP) and deep learning methods to automate the log statements' description prediction (RQ2). 
Our analysis shows that our hybrid NLP and code-clone detection approach (\textit{NLP CC'd}) outperforms conventional clone detectors in finding log statement \textit{locations} on average by 15.60\% and achieves 40.86\% higher performance on BLEU and ROUGE scores for predicting the \textit{description} of logging statements when compared to prior research (RQ3). 
Our work demonstrates the effectiveness of borrowing context from similar code snippets for automated log \textit{location} and \textit{description} prediction.
% and our results outperform the prior work.\looseness=-2 

\end{abstract}

%%
%% The code below is generated by the tool at http://dl.acm.org/ccs.cfm.
%% Please copy and paste the code instead of the example below.
%%

%%
%% Keywords. The author(s) should pick words that accurately describe
%% the work being presented. Separate the keywords with commas.
%\keywords{datasets, neural networks, gaze detection, text tagging}

%%
%% This command processes the author and affiliation and title
%% information and builds the first part of the formatted document.
\maketitle

\section{Introduction}
Developers embed logging statements into the software's source code to gather feedback on the computer systems' internal state, variables, and runtime behavior as a common practice. 
While the system is running, the output of logging statements is recorded in log files, which developers and system administrators will review at a later time for various purposes, such as anomaly and problem detection~\cite{xu2009detecting,fu2009execution}, log message clustering~\cite{makanju2009clustering,vaarandi2015logcluster}, system profile building, code quality assessment~\cite{shang2015studying}, and code coverage analysis~\cite{chen2018automated}. 
Additionally, the importance and depth of knowledge available in logs has also flourished the development of commercial log analysis platforms such as Splunk~\cite{urlsplunk} and Elastic Stack~\cite{urlelastic}. 
Figure~\ref{log_example} shows a log print statement (LPS) that contains a textual part indicating the context of the log, \textit{i.e.}, \textit{description}, a \textit{variable} part, and a log \textit{verbosity level} indicating the importance of the logging statement and how the level represents the state of the program. 
Verbosity levels for Log4j~\cite{log4x}, an Apache logging Library, include: \textit{trace, debug, info, warn, error, and fatal}.
\begin{figure}[h]
%\vspace*{-3mm}
%%\begin{flushleft}
\centering
\fbox{\begin{minipage}{24em}
log.warn(``Cannot find BPService for bpid=" + id);\\
\textbf{\hspace*{5mm}level}   $\vert$ \hspace{3mm}  \hspace{3mm} \textbf{description (LSD)} \hspace{13mm} $\vert$ \textbf{variable}
\end{minipage}}
%%\end{flushleft}
\caption{A log example with \textit{level}, \textit{description}, and \textit{variable} parts.}
\label{log_example}
\end{figure}

Due to the free-form text format of log statements and lack of a general guideline, adding proper logging statements to the source code remains a manual, inconsistent, and often an error-prone task~\cite{chen2017characterizing,chen2017characterizinganti}. 
In addition, in some cases developers forget to even add a log statement in the first place, \textit{i.e.}, missing log statements~\cite{hassani2018studying,li2021studying}. 
Moreover, because logging inherently introduces development and I/O cost~\cite{zhao2017log20,ding2015log2,gholamian2021distributed}, developers often struggle to decide the \textit{number}, \textit{location}, and \textit{description} of logging statements efficiently~\cite{jia2018smartlog}. 
Logging insufficiently may cause missing important runtime information that can negatively affect the postmortem dependability analysis~\cite{yuan2012conservative}, and unnecessary logging can consume extra system resources at runtime and impair the system's performance as logging is an I/O intensive task~\cite{zhao2017log20,ding2015log2,gholamian2021distributed}. 

Motivated with the aforementioned challenges, \textit{i.e., 1) ad hoc and forgetful logging practices, and 2) cost associated with inefficient logging}, prior methods aiming to automate logging \textit{\textbf{location}} and predict the \textbf{l}og \textbf{s}tatements' \textit{\textbf{description}} (LSD), \textit{i.e.}, the \textit{`static text'} of logging statement, are high in demand. 
These approaches generally aim to automate the logging process and predict whether or not a code snippet needs a logging statement by utilizing machine learning techniques to \textit{train} a model on a set of logged code snippets, and then \textit{test} it on a new set of unlogged code~\cite{fu2014developers,zhu2015learning} (supervised learning).  
However, prior research has come short in proposing a general solution based on an arbitrary code snippet, as they mostly concentrate on error log statements, such as \textit{exception} handling~\cite{zhu2015learning}. 
Another group of studies~\cite{he2018characterizing,gholamian2020logging} have shown the feasibility of predicting the log statements by the use of \textbf{\ul{similar code snippets}} or code clones\footnote{For code clones, we consider a wide range of similar code snippets, \textit{i.e.}, this includes copy-pasted code clones up to \ul{semi semantic clones} in the twilight zone~\cite{saini2018oreo}.}. 
Most recently, Gholamian~\cite{gholamian2021leveraginganonymous} produced a research plan and presented the motivation behind using code clone detection and NLP approaches to automate logging statements for new code snippets by borrowing the logging context from an already-existing similar code. 
In summary, the idea is that similar code snippets, \textit{i.e.}, clone pairs, follow similar logging patterns.

Accordingly, for introducing a general approach for log prediction and automation in our work, we study the application of similar code snippets and natural language processing (NLP) techniques for suggesting logging statements. 
In addition, our approach provides additional context to also predict other details of log statements, such as the LSD, which other log prediction approaches are unable to do. 
Initially, our study reveals that although using similar code snippets facilitates log automation, however, currently available tools for similar code detection (\textit{i.e.}, general-purpose clone detectors, such as Oreo~\cite{saini2018oreo}) cannot be used out-of-the-box for log statement automation. 
Thus, we first show that in order to enable log statement borrowing from similar code snippets, general-purpose code clone detectors require to be updated to understand logging statements. 
We perform this task by introducing LACCP, an improved log-aware clone-detector \textbf{(Finding 1)}. 
We then enable predicting the log statements' \textit{description} based on borrowing the logging statement from its clone pair and applying NLP and deep learning approaches \textbf{(Finding 3)}. 
Finally, we evaluate the effectiveness of our proposed approaches for log location and description automation with prior work and show that our approach outperforms conventional clone detectors in finding code snippets which require logging statement on average by 15.60\% and achieves 40.86\% higher performance on BLEU and ROUGE scores for predicting the \textit{description} of logging statements \textbf{(Findings 2 \& 4)}. 
In sum, our contributions in this research are as follows:
\begin{itemize}
\item We propose an \textbf{\ul{improved}} log-aware clone detection tool (\textit{LACCP}), which was initially introduced as LACC~\cite{gholamian2020logging} for log statements' \textit{`location'} prediction, by resolving two of the clone detection shortcomings (\S~\ref{shorfalls}).\looseness=-1
\item We introduce an algorithm to utilize \textit{LACCP} for LSD prediction and introduce a deep-learning NLP-based approach, ``\textit{(NLP CC'd)''}\footnote{The name resembles NLP-aware Code-Clone-based LSD suggestion.}, to work in collaboration with \textit{LACCP}, and to improve the performance of log statements' description prediction (\S~\ref{description}). 
We make our data available for both LACCP and NLP CC'd to encourage comparison and further research~\cite{rep_package_laccp}. 
\item We provide experimentation on several projects and measure Precision, Recall, F-Measure, and Balanced Accuracy, and compare LACCP's performance with general-purpose state-of-the-art clone detectors, Oreo~\cite{saini2018oreo} and LACC~\cite{gholamian2020logging}. 
In addition, we calculate the BLEU and ROUGE scores for our auto-generated log statements' \textit{descriptions} with considering different sequences of LSD tokens, and compare our performance with the prior work~\cite{he2018characterizing} (\S~\ref{rq3}).\looseness=-1 
\item We present a case study of the application of our tool for log description prediction in the real world and show how \textit{(NLP CC'd)} can facilitate the software development process (\S~\ref{casestudy}). 
\end{itemize}

The rest of this paper is organized as follows.
Section~\ref{defs} explains our motivation and methodology for log prediction with borrowing from similar code, and in Sections~\ref{shorfalls}, \ref{description}, and \ref{rq3}, we investigate RQ1, RQ2, and RQ3, respectively.  
We then provide a case study of the application of our approach in Section~\ref{casestudy}, and discuss the applicability of our approach in Section~\ref{discussion}. 
We provide the threats to the validity of our research in Section~\ref{threats} and review related work in Section~\ref{rwork}. 
At last, we present our conclusions and future directions in Section~\ref{conc}.

\section{Motivation and Methodology}\label{defs}
This section describes the motivation behind utilizing similar code snippets for log prediction followed by our methodology. 
\subsection{Motivation}
%\textbf{Motivation.} 
Improving logging quality with automated approaches is a crucial problem in software development as it helps to enhance the overall code quality~\cite{shang2015studying} and makes the system easier to debug~\cite{yuan2012conservative}. 
Thus, \textit{`where and what to log'} are the major challenges to tackle when developing tools to help developers with better logging. 
Prior research~\cite{gholamian2020logging,gholamian2021leveraginganonymous} shows that code clones resemble similar logging patterns and proposes an approach that utilizes them for log statement prediction by extracting features from method-level code blocks containing a logging statement.
\begin{figure}[h]
%\vspace*{-4mm}
\centering
\hspace*{-5mm}
\begin{minipage}{14cm}
%\hspace*{-3mm}
\begin{minipage}{0.55\linewidth}
\scriptsize
\hspace*{-5mm}
\begin{lstlisting}[xleftmargin=-2cm,numbersep=2pt,linebackgroundcolor={%
    \ifnum\value{lstnumber}=1
            \color{blue!10}
    \fi
    \ifnum\value{lstnumber}=9
            \color{gray!20}
    \fi
    },label={w_o_log_lvl_2},style=base]
//Original code - MD_i
int BS_recursive(A[], Key, l, h)
{
 if(l<=h)
 {  
  mid=(l+h)/2;
  if(Key==A[mid])
  {
   log.info("Found Key %d at Index %d.",Key,mid) 
   return mid;
  }
  else if(Key<A[mid])
   return BS_resursive(A[], Key, l, mid-1);
  else if(Key>A[mid])
   return BS_resursive(A[], Key, mid, h);      
 }
 return -1 //not found
}
\end{lstlisting}
\end{minipage}
\begin{minipage}{0.45\linewidth}
\scriptsize
\begin{lstlisting}[xleftmargin=-2cm,numbersep=2pt,linebackgroundcolor={%
    \ifnum\value{lstnumber}=1
            \color{blue!10}
    \fi
    },label={w_o_log_lvl_2},style=base]
//Clone Type 4 - MD_j
int BS_iterative(A[], Key, l, h)
{
 while (l<=h)
 {
  mid=(l+h)/2;
  if(Key==A[mid])
   return mid;
  else if (Key>A[mid]) 
   l=mid+1;
  else
   h=mid-1;
 }
 return -1; //not found
}
\end{lstlisting}
\end{minipage}
\end{minipage}
\caption{Log prediction with similar code snippets, \textit{i.e.}, semantic clones. On the left side, we observe the recursive psuedocode implementation of the binary search ($MD_i$), and on the right the iterative version ($MD_j$). Borrowing from similar code, the logging statement for $MD_j$ can be learned from its clone logging statement on Line 9 of $MD_i$.}
\label{code_sample}
\end{figure}

\subsection{Code Clones}
Similar code snippets (\textit{i.e.}, code clones) are code snippets that semantically are similar but can be syntactically different~\cite{saini2018oreo}. 
There are four main classes of code clones~\cite{rattan2013software}: Type-1, which is simply copy-pasting a code snippet, Type-2 and Type-3, which are clones that show syntax differences to some extent, and finally Type 4, which represents two code snippets that are semantic clones, \textit{i.e.}, they are syntactically very different but semantically equal~\cite{gholamian2021leveraginganonymous}. Figure~\ref{code_sample} shows the recursive versus iterative implementations of the \textit{binary search} (BS). 
The logging pattern in the original code, $MD_i$ on Line 9 can be learned to suggest logging statements for its clone, $MD_j$, which is missing a logging statement. 

Thus, to predict log statements, we define relevant log-aware source code features and employ them for predicting whether a newly composed method code block requires a log statement. 
We use \ul{method-level clones} (rationale explained in the following) and apply different categories of source-code features and feed them into a machine learning tool to identify similar code snippets (\textit{i.e.}, clone pairs), which also follow similar logging patterns. 
Formally speaking, assuming set $CC_{MD_i}$ is the set of all code clones of Method Definition $MD_i$, if $MD_i$ has a log print statement (LPS), then its clones also have LPSs:

\begin{minipage}{\linewidth}
\centering
{\hlcyan{$\exists LPS_i\in MD_i \implies \forall MD_{j}\in CC_{MD_i}, \exists LPS_j \in MD_{j}$}.}
\end{minipage}

\subsection{Why Leveraging Code Clones for Log Prediction?} 
We observe the benefits of using our approach in utilizing similar code snippets to borrow logging patterns are threefold:
\begin{enumerate}[label=\protect\circled{\arabic*}]
\item Clone detection methods are already a part of the software maintenance process. 
Therefore, it is beneficial if we rely on approaches that already exist in the development process. It enables the reuse of stable tools and techniques, saves on development cost, and expedites the process.
\item Although we acknowledge that in some cases, code clones are the outcome of shallow copy-pasting which results in log-related
anti-patterns (\textit{i.e.}, issues)~\cite{li2021studying}, this, simultaneously, shows the potential of code clones as a starting point for automated log suggestion and improvement. In other words, by automating and enhancing the log statements in the clone pairs, we can expedite the development process and avoid shallow copy-pasting that
developers tend to do. Additionally, by automation, we reduce the risk of irregular and \textit{ad hoc} developers' logging practices, \textit{e.g.}, forgetting to log in the first place. 
\item A significant amount of research is conducted towards improving clone detection in identifying semantic clones~\cite{gharehyazie2019cross,yu2019neural,wang2020detecting}. 
Thus, we foresee our approach becomes emboldened as clone detection approaches grow to be more elaborate, and code reuse and context borrowing will further facilitate and expedite the software development process~\cite{gharehyazie2019cross}.  
\end{enumerate}

\subsection{Method-Level Log Prediction Rationale}
In our approach, we decide whether a method code block requires a logging statement, \textit{i.e.}, method-level log decisions. Although finding similar code snippets, and subsequently, log statement prediction can be performed in different granularity levels, such as files, classes, methods, and code blocks, however, method-level clones appear to be the most favorable points of re-factoring for all clone types~\cite{kodhai2014method,gholamian2021leveraginganonymous}. 
This approach also includes all of the logging statements which are nested inside more preliminary code blocks within method definitions, \textit{viz.}, logging statements nested inside code blocks, such as \textit{if-else} and \textit{try-catch}. 
We also hypothesize that the core idea of our research, \textit{i.e.}, context borrowing from similar code snippets, can be extended to an arbitrary code snippet without major changes.

\subsection{Practical Scenario}
A practical scenario that showcases the usability of our approach for log statement location and description during the software development cycle is as follows. 
We consider a possible employment of our research as a recommender tool, which can be integrated as a plugin to code development environments, \textit{i.e.}, IDE software. 
Alex is a developer working on a large-scale software system, and he has previously developed method $MD_i$ in the code base. 
At a later time, Sarah, Alex's colleague, is implementing $MD_j$, which does not have a logging statement yet. 
Our automated log suggestion\footnote{We use \textit{`automation'}, \textit{`suggestion'}, and \textit{`prediction'} interchangeably.} approach can predict that if this new code snippet, $MD_j$, requires a logging statement by finding its clone, $MD_i$, in the code base. 
Then, the tool can suggest Sarah, just in time, to add a logging statement based on the prediction outcome.
Although prior work has shown the majority of the clone pairs match in their logging behavior~\cite{gholamian2020logging}, when there is a conflict among clones, we can return a list of suggestions to the developer and in the end, the developer will make the final decision. 
This approach can likewise recommend the logging statement description by retrieving the text existing in its clone log statement and the description predicted by the NLP model. 
This situation can be extended to several source code projects, and bring intra- and inter-project logging suggestions. 
The latter is useful for small projects where there is not a sufficient prior code base, which is commonly referred to as the project's ``cold start'' phase~\cite{guo2016cold}.

\subsection{Research Questions}
We guide our study with the following research questions (RQs):
\begin{enumerate}[label=\textbf{(RQ\arabic*)}]
    \item How code clones can be used for automated log location prediction?
    \item How the available context from clone pairs can be borrowed for log description prediction?
    \item How the accuracy of both log location and description prediction can be evaluated and compared with prior work? 
\end{enumerate}

For RQ1, we first expose two shortcomings of general-purpose clone detection, and then improve on the performance of clone detection methods to make them more suitable for effective log statement \textit{location} prediction (\S~\ref{shorfalls}). 
For RQ2, we predict the \textit{description} of logging statements in methods without logging statements by searching for the logging descriptions that we can obtain from their clone pairs. Additionally, to enhance the LSD prediction, we apply NLP-based deep learning (DL) methods (\textit{NLP CC'd}) and further improve the LSD prediction when compared to the LSD retrieved from the clone pair (\S~\ref{description}). 
Finally, we evaluate the performance of both log location and description prediction in RQ3 (\S~\ref{rq3}).

\section{ RQ1: How code clones can be used for automated log location prediction?}\label{shorfalls}
\subsection{Motivation and Approach} Prior work has shown similar code snippets have similar logging characteristics~\cite{he2018characterizing,gholamian2020logging,gholamian2021leveraginganonymous}. 
This finding opens up a potential way to automate logging statements' locations~\cite{gholamian2021leveraginganonymous}. 
The approach in essence is to find similar code snippets to the code that is currently being developed, and make a logging decision for the new code by observing the logging patterns of its similar code. 
This automated log suggestion approach can help developers in making logging decisions and improve logging practices.\looseness=-1 

\subsection{Findings} Although the proposed approach has potential for log automation, during our initial manual scrutiny of detected and undetected similar code snippets, we observe that due to \textbf{two shortcomings} that exist in the prior work, we have not been able to gain the full benefit of the log automation through similar code. 
In particular, we discover that although prior work~\cite{gholamian2020logging} observed satisfactory prediction scores for the projects under study with LACC and outperforms Oreo~\cite{saini2018oreo}, it falls short in balancing the Precision and Recall of predictions. 
As such, we hypothesize that the log-prediction performance of LACC can be improved by recognizing some of the \textit{`not-detected (false negative)'} and \textit{`mis-detected (false positive)'} cases of clone detection, which are directly pertinent to the existence or absence of the logging statements. 
In the following, we assume $(MD_i, MD_j)$ are methods with logging statements that initially detected as clone pairs, and the prime-symbol version ($'$), \textit{e.g.}, $MD_{j'}$, is obtained from the method after removing its logging statement.\looseness=-1 

\subsubsection{Shortcoming I (SI) - high rate of not-detected clones.}
This scenario happens when the logging statement(s) from the method definition in the code base ($MD_i$) contribute to a significant portion of the method body. $(MD_i, MD_j)$ are detected as clone pairs, because both of them have a logging statement and thus their code feature values match. 
However, in a real-world scenario, when $MD_{j'}$ is being just developed and does not have a logging statement yet, the method's source code metrics for $MD_{j'}$ will be significantly different from $MD_i$, which results in $MD_i$ and $MD_{j'}$ to not be detected as clone pairs \textit{(i.e.}, false negative)~\cite{gholamian2020logging}. 

\subsubsection{Shortcoming II (SII) - high rate of mis-detected clones.}\label{short2} Because log statements do not change the semantics of the source code, we argue that two code snippets should be clones regardless of the existence of their log statements. 
We notice there are a considerable number of code clones that are not matched as clone pairs after the log statements (including the log statement static text) are removed from both clone pairs, $MD_{i'}$ and $MD_{j'}$, \textit{i.e.}, false positive. 
In other words, log statements had a critical role in matching the clone pairs, and if log statements are removed from both code snippets, then the code snippets are no longer detected as clone pairs. 
Considering that log statements are for book-keeping purposes, and do not change the semantics of the program, we reckon this case as a false positive clone detection that only relies on the similarities of log statements rather than the \textit{semantic-effective} lines of the source code. 
Listing~\ref{wlog_w_o_logs} illustrates this case as $MD_i$ and $MD_j$ are only matched because of the similarity in their logging statements, which is a  book-keeping aspect and does not change the semantics of the code. 

\begin{figure}[h]
%\vspace{-5mm}
\begin{minipage}{\linewidth}
%\belowcaptionskip=-10pt
\begin{lstlisting}[linebackgroundcolor={%
    \ifnum\value{lstnumber}=1
            \color{blue!10}
    \fi
    \ifnum\value{lstnumber}=15
            \color{blue!10}
    \fi    
    },
caption={Wrong clone detection because of log statements. },label={wlog_w_o_logs},style=base]
//MD_i, logging statements are commented. 
  protected byte[] createPassword(NMTokenIdentifier identifier) {
    //LOG.debug("creating password for {} for user {} to run on NM {}",
    //   identifier.getApplicationAttemptId(),
    //    identifier.getApplicationSubmitter(), identifier.getNodeId());
    readLock.lock();
    try {
      return createPassword(identifier.getBytes(),
          currentMasterKey.getSecretKey());
    } finally {
      readLock.unlock();
    }
  }
  
//MD_j, logging statements are commented. 
protected byte[] retrivePasswordInternal(NMTokenIdentifier identifier,
      MasterKeyData masterKey) {
    //LOG.debug("retriving password for {} for user {} to run on NM {}",
    //    identifier.getApplicationAttemptId(),
    //    identifier.getApplicationSubmitter(), identifier.getNodeId());
    //LOG.debug("Response line: " + identifier.getResponseLine());
    return createPassword(identifier.getBytes(), masterKey.getSecretKey());
  }
\end{lstlisting}
\end{minipage}
%\vspace{-4mm}
\end{figure}

\subsubsection{Overcome the Shortcomings}
The observations from \textbf{SI} and \textbf{SII} confirm that general-purpose clone detection cannot readily be applied for log suggestion, as we are looking to suggest a log statement for a newly-developed code snippet (\textit{i.e.}, without a logging statement) by finding its clone pairs that already have logging statements. 
As such, with log-aware feature calculation in LACCP, we aim to achieve a higher performance in clone-based log statement automation. 

Table~\ref{table_metrics} presents log-related features which are utilized for detecting method clones with a logging statement. 
These features are in three main categories: \textit{numeric, boolean, and string}. 
For example, \textit{LWK} represents log related keywords and wrappers, \textit{e.g.}, `log.info' and `logger.debug' as string features. 
The selected features in Table~\ref{table_metrics} enable us to recognize the logging statements and consider them respectively in source code feature calculation. 
We have surveyed the features used in prior work~\cite{samoladas2008sqo,saini2018oreo} and experimented with them, \textit{i.e.}, with feature selection and extraction~\cite{khalid2014survey}, and measured the performance metrics such as \textit{Precision} and \textit{Recall}, and picked the features which help the most with log prediction accuracy.\looseness=-1
{\renewcommand{\arraystretch}{1}
\begin{table}[h]
\footnotesize
%\vspace*{-3mm}
\centering
\begin{tabular}{|p{2cm} |p{6cm} |p{2cm} | } 
 \toprule
\rowcolor{blue!10} \textbf{Feature} & \textbf{Description} & \textbf{Type}  \\ 
 \midrule
  ELPS & Existence of an LPS & Boolean \\
\midrule
 NTOK & Number of tokens & Numerical\\
\midrule
NOS & Number of statements& Numerical  \\
\midrule
NEXP & Number of expressions& Numerical \\
\midrule
LMET & Number of local methods called & Numerical \\
\midrule
XMET & Number of external methods called & Numerical \\
\midrule
SLOC & Source lines of code& Numerical \\ 
\midrule
 LWK & Logging wrappers and keywords & String \\ 
\bottomrule
\end{tabular}
\vspace{.5mm}
\caption{Method-level log related features.}
\label{table_metrics}
\vspace*{-4mm}
\end{table}
}

\subsection{Log-Aware Feature Calculation Illustrative Example}
To elaborate further on the inner workings of LACCP, we provide the following example. 
The idea is to examine how the method-level features from Table~\ref{table_metrics} are calculated with and without LACCP. 
In Table~\ref{table_metrics}, \textit{SLOC} includes all lines of the source code, such as comments, brackets (\{\}) for \textit{if-else} blocks, \textit{etc.}; however, \textit{NOS} includes only executable expressions in the method definition. 
Following examples are based on Line 21: \textit{``LOG.debug("Response line: " + identifier.getResponseLine())''} from Listing~\ref{wlog_w_o_logs}.

\textbf{ELPS:} there exists logging statements for this method, therefore, $ELPS_{MD_i}=True$. 

\textbf{NTOK:} in Listing~\ref{wlog_w_o_logs}, removing this line causes the number of tokens for $MD_{j'}$ (\textit{i.e.}, $NTOK_{MD_{j'}}$) to decrease by 6 when compared to $NTOK_{MD_i}$; tokens are \textit{`LOG'}, \textit{`debug'}, \textit{`Response'}, \textit{`line'}, \textit{`identifier'}, and \textit{`getResponseLine'}. 

\textbf{SLOC, NOS, and NEXP:} similar to NTOK, $SLOC_{MD_{j'}}$, $NOS_{MD_{j'}}$, and $NEXP_{MD_{j'}}$ values reduce by one as an executable line of $MD_{j'}$ has been removed.

\textbf{LMET and XMET:} these values represent the number of local and external method calls. 
$LMET_{MD_{j'}}$ and $XMET_{MD_{j'}}$ values also decrease as the omitted line makes references to other methods, both internal, \textit{`getResponseline'}, and external, \textit{`debug'}. 

\textbf{LWK:} we also search and find a comprehensive set of log related keywords, \textit{e.g.}, `log.info', `logger.debug', \textit{etc.}, as string features, which come into consideration in LACCP. 
Table~\ref{log_aware_ilus_example} summarizes the changes in feature values for $MD_{j}$ and $MD_{j'}$. 

{\renewcommand{\arraystretch}{.6}
\begin{table}[h]
\vspace*{1mm}
\footnotesize
\centering
\begin{tabular}{ | p{2cm} | p{2.4cm} | p{3cm}|} 
\toprule
\rowcolor{blue!10}Feature &  Value ($MD_{j}$) & Value ($MD_{j'}$) \\ 
\midrule
ELPS& True & Flase\\ 
\midrule
NTOK& $NTOK_{MD_{j}}$ & $NTOK_{MD_{j}}-6$\\ 
\midrule
SLOC& $SLOC_{MD_{j}}$ & $SLOC_{MD_{j}}-1$\\ 
\midrule
NOS& $NOS_{MD_{j}}$ & $NOS_{MD_{j}}-1$\\ 
\midrule
NEXP& $NEXP_{MD_{j}}$ & $NEXP_{MD_{j}}-1$\\ 
\midrule
LMET& $LMET_{MD_{j}}$ & $LMET_{MD_{j}}-1$\\ 
\midrule
XMET& $XMET_{MD_{j}}$ & $XMET_{MD_{j}}-1$\\ 
\midrule
LWK& $LOG.debug$ & $None$\\ 
\bottomrule
\end{tabular}
\caption{Log-related features comparison with ($MD_{j}$) and without ($MD_{j'}$) the log statement.}
\label{log_aware_ilus_example}
%\vspace{-4mm}
\end{table}
} 

In a real scenario, $MD_i$ is previously developed and is in the code base, and we are looking to automate logging for its clones which are being currently developed without logging statements (\textit{i.e.}, $MD_{j'}$). 
Thus, the more logging statements exist in method $MD_i$, the more source code features will diverge for $MD_i$ and $MD_{j'}$, and thus it becomes more troublesome for general-purpose clone detectors to detect them as clone pairs. 
In LACCP's design, for each method $MD_i$ with a logging statement, we calculate the features by recognizing the logging code first and then exclude its impact on the values of the features in Table~\ref{table_metrics}. 
Feature values are updated such that methods are detected as clone pairs regardless of the presence of logging statements, \textit{i.e.}:

{\vspace{1mm}
\hlcyan{$Fr(MD_i)\sim_{clone} Fr(MD_j) \implies Fr_{LACCP}(MD_i)\sim Fr(MD_{j'}) \implies Fr_{LACCP} (MD_{i'}) \sim Fr(MD_{j'})$}
\vspace{1mm}}.

We then add the methods which satisfy the above condition to the set of clone pairs for $MD_i$. 
In addition to the features in Table~\ref{table_metrics}, we also utilize the other features listed in~\cite{saini2018oreo} for general clone detection, however, we only perform \textit{\textbf{log-aware}} feature calculation on features in Table~\ref{table_metrics}. 
Since log statements do not directly change other feature values, we refer the reader to~\cite{saini2018oreo} for further details. 
An example of features that log statements do not generally have impact on is the number of loops, \textit{i.e.}, \textit{for and while}.

\subsection{Approach Significance} 
Prior approaches~\cite{zhu2015learning,li2020shall} rely on extracting features and training a learning model on logged and unlogged code snippets. 
Thus, they can predict if a new unlogged code snippet needs a logging statement by mapping its features to the learned ones. 
Although these methods initially appear similar to our approach in extracting log-aware features from code snippets, \textit{i.e.}, Table~\ref{table_metrics}, we believe our approach has an edge over the prior work. 
Because we also have access to the clone pair of the code under development, \textit{i.e.}, $MD_i$ in $(MD_i, MD_{j'})$, this enables us to obtain and borrow the additional context from $MD_i$ to predict other aspects of log statements, \textit{e.g.}, the LSD, which the prior work is unable to do. 
The significance of our approach becomes apparent in LSD automation (\S\ref{description}) as we utilize the LSD of the clone pair as a starting point for suggesting the LSD of the new code snippet. 
Thus, our approach not only complements the prior work in providing logging suggestions for developers as they develop new code snippets, but it also has an edge over them by providing additional context for further prediction of LPS details, such as the LSD and the log's verbosity level. 
Moreover, prior research has shown~\cite{gholamian2020logging} that there exists a significant portion of clone pairs of Type 3 and above, \textit{i.e.}, code pairs that are considerably different in syntax or they are semantic pairs. 
Although later on we evaluate and show the applicability of our approach on a set of limited projects, we envision that our approach would be of a greater significance for a large collection of software, \textit{e.g.}, thousands of projects from GitHub. 
This way, semantic clone pairs can be found across different projects and used to borrow and predict log statements.

\section{RQ2: how the available context from clone pairs can be borrowed for log description prediction?}\label{description}
\subsection{Motivation} Based on the approach for predicting the location of logging statements with similar code snippets in RQ1 and the additional available context from the clone pairs, \textit{i.e.}, the logging statement description available from the original method, $MD_i$, we hypothesize it is a valuable research effort to explore whether it is also possible to predict the logging statements' \textit{description} automatically. 
With satisfactory performance, an automated tool that can predict the description of logging statements will be a great aid, as it can expedite the logging process and improve the quality of logging descriptions.

\subsection{NLP for LSD Prediction - Theory}
%\textbf{NLP for LSD prediction.} 
The predictable and repetitive characteristics of common English text, which can be extracted and modeled with statistical natural language processing (NLP) techniques, have been the driving force of various successful tasks, such as speech recognition~\cite{bahdanau2016end} and machine translation~\cite{marino2006n}. 
Prior research~\cite{hindle2012naturalness,tu2014localness,allamanis2014mining,gholamian2021naturalness} has shown that software systems are even more predictable and repetitive than common English, and language models perform better on software engineering tasks than English text; tasks such as code completion~\cite{raychev2014code} and suggestion~\cite{bhoopchand2016learning}. 
Most recently, He \textit{et al.}~\cite{he2018characterizing} and Gholamian and Ward~\cite{gholamian2021naturalness} showed that logging descriptions in the source code and log files also follow natural language characteristics. 
Thus, we introduce a deep learning (DL) natural language model to auto-generate the log statements descriptions. 
Intuitively, if there is observable repetitiveness in logging descriptions, the trained model should have acceptable prediction performance for new logging statements.

There are two main categories of language models (LMs): 1) statistical LMs which utilize n-gram~\cite{ngrammodel} and Markovian distribution~\cite{urlmarkov} to learn the probability distribution of words, and more recently, 2) deep learning (DL) LMs which have surpassed the statistical LMs in their prediction performance, as they can capture more long-range token dependencies~\cite{white2015toward,das2015contextual}. 
Thus, in this research, we utilize deep learning LMs. 
Once LMs are trained on sequences of tokens or n-grams (\textit{e.g.}, words), they can assign scores and predict the probability of new sequences of words. 
Considering a sequence of tokens in a text (in our case, logging statement description, LSD), $S=a_1,a_2, . . ., a_N$, the LM statistically estimates how likely a token is to follow the preceding tokens. 
Thus, the probability of the sequence is estimated based on the product of a series of conditional probabilities~\cite{hindle2012naturalness}:
\begin{equation*}\label{p_theta_s}
P_{\theta}(S) =  P_{\theta}(a_1)P_{\theta}(a_2|a_1)P_{\theta}(a_3|a_1a_2)....P_{\theta}(a_N|a_1...a_{N-1}) 
\end{equation*}
which is equal to:\vspace*{-2mm}
\begin{equation}\label{p_theta_sum}\vspace{-2mm}
P_{\theta}(S) =  P_{\theta}(a_1) . \prod_{t=2}^{N} P_{\theta} (a_{t}| \vspace*{2mm}a_{t-1}, a_{t-2}, ..., a_1),
\end{equation} 
where $a_1$ to $a_N$ are tokens of the sequence S and the distribution of $\theta$ is estimated from the training set. 
Given a sequence of log description tokens $a_1,..., a_t$, we seek to predict the next M tokens $a_{t+1}, ..., a_{t+M}$ that maximize Equation~\ref{p_theta_sum}~\cite{bhoopchand2016learning}:

\begin{equation}\label{max_p}
P_{\theta}(S) =   \underset{a_{t+1}, ..., a_{t+M}} {arg \hspace{1mm} max} P_{\theta} (a_1, ..., a_i, a_{i+1},..., a_{i+M})
\end{equation}

As such, an LSTM implementation of the LM to maximize the probability of observing token $a_t$ in Equation~\ref{max_p} at time step t is: 
\vspace*{-1mm}
\begin{equation}\label{lstm_formula}
P_{\theta}(a_t| a_{t-1}, ...,a_1) = \frac{exp(\upsilon_{a_t}^T h_t + b_{a_t})}{\sum_{a_{t'}} exp(\upsilon_{a_{t'}}^T h_t + b_{a_{t'}})}   
\end{equation}
where $h_t$ is the output of the hidden state vector at time t, $\upsilon_{a_t}^T$ is a parameter vector associated with token $a_t$ in the vocabulary and $b_{a_t}$ is a constant value. 
Intuitively, in Equation~\ref{lstm_formula}, $\upsilon_{a_t}^T h_t + b_{a_t}$ is a function that shows how much the model favors in observing $a_t$ after the sequence of $a_{t-1}, ...,a_1$, and the \textit{exp} function assures the values are always positive. 
The summation in the denominator calculates the probability values of each token over all tokens out of the maximum probability value of 1. 

\subsection{Methodology} 
We base our method on the assumption that clone pairs tend to have similar logging statements' descriptions. 
This assumption comes from the observations in predicting log statements for clone pairs. 
As logging descriptions explain the source code surrounding them, it is intuitive for similar code snippets to have comparable logging descriptions. 
Based on this assumption, we propose a deep learning method that borrows the LSD from similar code snippets and leverages NLP approaches (NLP CC'd). 
In particular, to generate the LSD for a logging statement in $MD_j$, we extract its corresponding code snippet (\textit{i.e.}, the method without the log statement, $MD_{j'}$) and leverage LACCP to locate its clone pairs. 
Laterally, the NLP model is trained on the logging descriptions available in the training set for each project. 
To ensure the training and testing sets are mutually exclusive, for all of the clone pairs of $(MD_i, MD_j)$, the LSDs of $MD_i$s and $MD_j$s create the training and testing data, respectively.
During the testing, the retrieved logging description from the clone pair ($LSD_{MD_i}$) is served as a starting input point for the NLP model to propose a set of description suggestions. 
Then, we evaluate the similarity between the NLP-generated descriptions with the LSD provided by the developers ($LSD_{MD_j}$) (\textit{ground truth}). 
Our methodology resembles the scenario that while the developer is creating a new snippet of the source code, we look for its similar code snippets with LACCP, and in case a clone is found ($MD_i$), we work further to provide predictions on the description of the logging statement by generating suggestions from the available clone's LSD ($LSD_{MD_i}$) combined with the collective knowledge of LSDs available in the code base.

\begin{figure}[t]
 \centering
\includegraphics[width=0.9\textwidth]{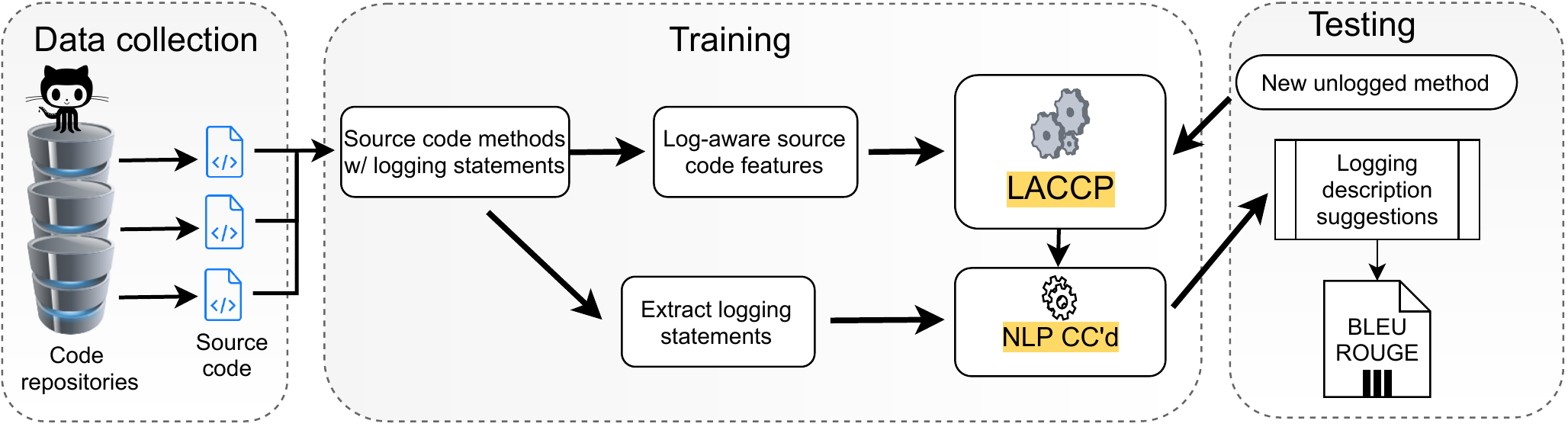}
\caption{The toolchain for log statement description prediction. The approach shows how both LACCP and NLP CC'd collaborate for LSD prediction.}
\label{toolchain}
%\vspace*{-2mm}
\end{figure}

\subsection{Toolchain}
Figure~\ref{toolchain} presents our toolchain for log statement description prediction. 
In the \textbf{data collection} phase, we select open-source Java projects from their Git repositories, based on factors of interest such as age and size of the project (in source lines of code), popularity (being well-established), stability, and logging index of the projecst~\cite{chen2017characterizing}. 
Listed in Table~\ref{systems}, we select seven Apache Java projects. %Hadoop, Zookeeper, HBase, and CloudStack. 
Next, commencing in the \textbf{training} phase, we extract method definitions (MDs) containing logging statements by applying JavaParser~\cite{jparser}. 
Initially, we parse the source code to obtain the abstract syntax tree (AST), which is the hierarchical representation of the code. 
We use the AST to access Java method definitions with logging statements. 
We then extract method-level code features to perform log-aware clone detection (LACCP) on the extracted method definitions and leverage LACCP to find clone pairs with logging statements. 
Next, for each detected clone pair, we use the descriptions obtained from $MD_i$s to train the NLP model. 
Finally, in the \textbf{testing} phase, we use the $MD_{j'}$s as test-case inputs to LACCP. 
The NLP model, upon the clone pair detection of $(MD_i, MD_{j'})$, receives the LSD from $MD_i$ and suggests descriptions with the highest probability for $MD_{j'}$. 
Then, we compare the NLP-generated LSDs with the logging statement originally placed by developers in $MD_{j}$, and calculate the BLEU and ROUGE scores. 

\subsection{Implementation}
\begin{figure}[t]
\centering
\includegraphics[width=0.9\textwidth]{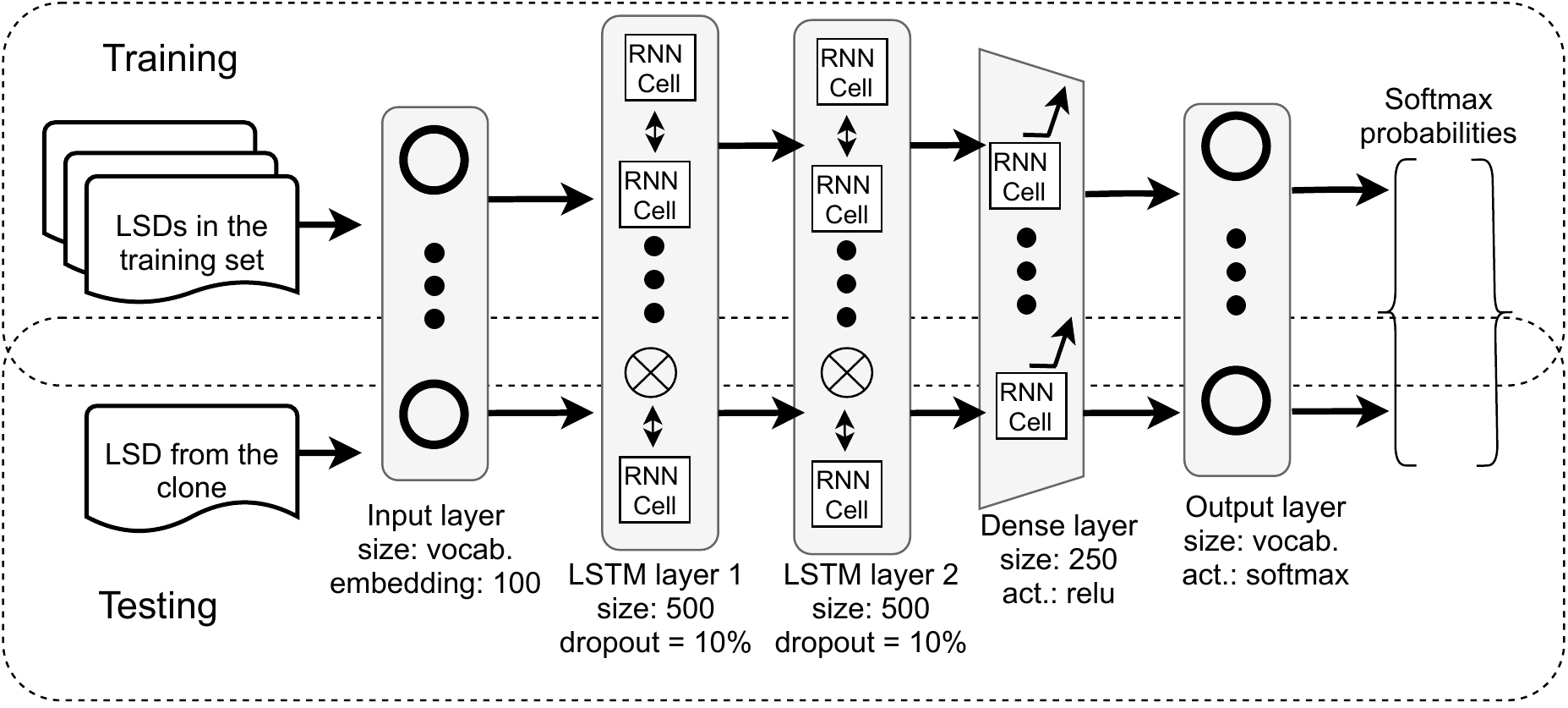}
\caption{The figure shows the inside of NLP CC'd, our deep learning long short-term memory (LSTM) model for log description prediction.}
\label{lsd_nlp}
\vspace{3mm}
\end{figure}

For our NLP approach, we utilize Long Short-Term Memory (LSTM) \cite{hochreiter1997long} models which are recurrent neural networks (RNNs) capable of capturing long-term dependencies in a sequence of tokens through their internal memory. 
This feature makes them suitable for LSD prediction in our research, as we are pursuing to predict a sequence of words for the LSDs. 
Figure~\ref{lsd_nlp} shows the overall layout of our model, which has an input layer, two hidden layers, a dense layer with Rectified Linear Unit (relu)~\cite{nair2010rectified,ramachandran2017searching} as the activation function, and an output layer with \textit{softmax} activation. 
The layers are sized as: input layer is the \textit{`vocabulary size'}, the LSTM layers 1 and 2 are \textit{`500 cells'}, the dense layer is \textit{`250 cells'}, and the output layer has the same size as the input layer. 
During the training phase, in the first layer of the model, we map the LSD sequences to vectors of integers by leveraging an \textit{`embedding layer'}. 
The embedding layer infers the relationships among tokens in the LSD input sequences, and outputs a set of lower-dimension vectors, called word embeddings~\cite{mikolov2013efficient}. 
The embedded vectors then pass through two layers of LSTM and allow the model to learn the relationship between the sequence of words in the LSD and assign probabilities. 
Followed by LSTM layers 1 and 2, the dense layer is placed with \textit{relu} activation function. 
Finally, the output layer produces \textit{softmax} probabilities for each next token prediction in the suggested LSD.

For the DL implementation, we use Python's Keras library~\cite{urlkeras}. 
We utilize a \textit{softmax} activation function~\cite{bridle1990probabilistic,xu2000can} on the output layer such that the network can learn and output probability distribution over possible next tokens in the sequence of words within the LSD. 
This ensures that the LSTM outputs are all in the range of [0,1], and their summation is equal to one in every prediction~\cite{graves2005framewise}. 
We also apply \%10 dropout on the hidden layers to avoid overfitting~\cite{srivastava2014dropout}. 
We train the model for 200 epochs and set the batch size to 64. 
During the testing, from the outputted LSDs of the DL model, we pick the highest (NLP-1) or top-3 (NLP-3) softmax probabilities and provide them as suggestions.  
We have been partially deliberate in the selection of the hyperparameters, and as an avenue for future work, different layouts or hyperparameters' setup, such as more memory cells or deeper layers of LSTM network, may achieve a better performance.

{\renewcommand{\arraystretch}{1}
\scriptsize
\begin{algorithm}[t]
\caption{\fontsize{7}{4.5}\selectfont \textbf{Log Location \& Description Predictor (LACCP \& NLP CC'd)}}\label{algo2}
\KwInput{Java source code repositories}
\KwOutput{$BLEU_{list}$ and $ROUGE_{list}$ scores}
 $sourceCode_{AST} \gets Parse(sourceCode)$ \label{ast_parse}
 
 $Methods_{w/LPS} \gets extract_{MD}(sourceCode_{AST}, exist({LPS}));$\label{wlsp}
 
  $nlp_{training} \gets \{\};$ $nlp_{testing} \gets \{\};$
 
 \For{\texttt{($\forall MD_i \in Methods_{w/LPS}$)}}
 {\label{f1}
         $CC_{MD_i}\gets$ \texttt{findClones$_{LACCP}$($MD_{i}$)}
         
        $CC\_pairs({MD_{i},MD_j})\gets createPairs(CC_{MD_i})$\label{createPairs}
        
        \For{\texttt{($\forall (MD_i,MD_j)\in CC\_pairs({MD_{i},MD_j})$)}}
 {\label{f2}
        $nlp_{training}$ $\gets$ $MD_i \bigcup$ $nlp_{training};$\label{addtonlp_train} 
         
          $nlp_{testing}$ $\gets$ $MD_{j'}  \bigcup$ $nlp_{testing};$   \label{addtonlp_train2}           
         
}\label{f1end}

}\label{f2end}

$BLEU_{list} \gets \{\};$

$ROUGE_{list} \gets \{\};$

$nlp_{training}  \gets preProcess(nlp_{training});$\label{pre_process}

$nlpModel  \gets train(\forall MD_i \in nlp_{training});$\label{train_nlp}

\For{\texttt{($\forall (MD_i, MD_{j'})| MD_{j'}\in nlp_{testing}$)}}
{\label{f22}
      $nlpLSD \gets nlpPredic(LSP(MD_i));$
      
       \For{\texttt{$\forall LSD_i \in nlpLSD$}}
       {        
         $BLEU_{list} \gets calcBLEUScores( LSD_i, LPS(MD_j));$

         $ROUGE_{list} \gets calcROUGEScores(LSD_i, LPS(MD_j));$
        }
          
}\label{f22end}

\textbf{return} $BLEU_{list}, ROUGE_{list};$\label{return2} 
\end{algorithm}
}

\subsection{LSD Prediction Algorithm and Steps}
%\textbf{LSD prediction algorithm and steps}
We follow the steps outlined in Algorithm~\ref{algo2} for LSD suggestion for a method based on the LSD borrowed from its clone pair. 
After collecting the source code projects from their git repositories, we then parse the source code to its abstract syntax tree (AST)~\cite{ast}, and search for methods with logging statements (Lines~\ref{ast_parse}-\ref{wlsp}). 
We define $CC_{MD_{i}}$ as the set of all clone pairs of Method $MD_i$, \textit{i.e.},: $ CC_{MD_{i}}(pairs) = \{ (MD_i,MD_j) | \forall MD_{j} \in CC_{MD_i} \} $. 
On Line~\ref{wlsp}, after extracting all method definitions with at least one LPS from the AST, we find all clone pairs for each method definition $MD_i$, by applying \textit{LACCP}, in the \textit{for-loop} on Lines~\ref{f1}-\ref{f1end}. 
After finding clones of $MD_i$ and creating $(MD_i,MD_j)$ pairs on Line~\ref{createPairs}, we add $MD_i$s to the training and $MD_{j'}$s to the testing sets on Lines~\ref{addtonlp_train}-\ref{addtonlp_train2} for NLP CC'd to use later for LSD prediction. 
We also add the pair to the sets of $nlp_{training}$ and $nlp_{testing}$ on Lines~\ref{addtonlp_train} and~\ref{addtonlp_train2}. 
Before training the LSTM model on the retrieved LSD data, we perform \textbf{pre-processing} on the LSDs (Line~\ref{pre_process}) such as tokenization, mapping punctuation to the vocabulary space, cleaning of special characters (\textit{\textit{e.g.}}, Unicode) and removing the LSD parts related to the dynamic variables. 
On Line~\ref{train_nlp}, we train the NLP model by using the logging descriptions collected in the training set. 
Then, in the \textit{for-loop} on Lines~\ref{f22}-\ref{f22end}, for predicting the LSD for $MD_{j'}$, we feed in the logging description from $MD_i$ to the NLP model, and the trained model returns LSD suggestions.%

\section{RQ3: how the accuracy of both log location and description prediction can be evaluated and compared with prior work?}\label{rq3}
\vspace*{-1mm}
In this section, we evaluate the performance of LACCP (\textbf{RQ3.I}) and NLP CC'd (\textbf{RQ3.II}).

{\renewcommand{\arraystretch}{1}
\begin{table}[h]
\vspace{2mm}
\scriptsize
\centering
\begin{tabular}{ | p{1.7cm} | p{3.4cm} | p{0.7cm}| p{0.7cm}| p{1.9cm}|} 
\toprule
\rowcolor{blue!10}Project (abrv.) &  Description &\# LOC  & \# LPS & \# of LPS per KLOC\\ 
\midrule

Hadoop (HD)& Distributed Computing& 2.10M& 16,202     & 7.72\\ 

\midrule
Zookeeper (ZK) & Configuration management& 94,434      & 1,885 & 19.96\\ 
\midrule
CloudStack (CS) &Cloud deployment & 739,732    & 12,237 & 16.54\\ 
\midrule
HBase (HB)& Distributed database &949,310   & 9,264 &9.76\\ 
\midrule
Hive (HI)& Data warehouse& 1.84M  & 10,640  & 5.78\\ 
\midrule
Camel (CL)& Integration platform & 2.2M  & 9,682 & 4.40\\ 
\midrule
ActiveMQ (MQ)& Message broker&  464,632  & 6,442 & 13.86\\ 
\bottomrule

\end{tabular}
\caption{The table lists the details for the studied project. The projects are well-established software from different application domains. The table also lists the number of lines of code (LOC), number of log print statements (LPS), and number of log statements per thousand lines of code (KLOC).}
\label{systems}
%\vspace{-4mm}
\end{table}
}

\subsection{Systems Under Study}\label{sys_under_study} 
Based on the research on open-source projects by Chen and Jiang~\cite{chen2017characterizing} and He \textit{et al.}~\cite{he2018characterizing}, we choose seven open-source Java projects. 
These projects are well-logged, stable, and well-used in the software engineering community, and this selection also enables us to compare our research with prior work, accordingly. 
Table~\ref{systems} summarizes the line number of source code (LOC), the number of logging statements (LPS) in each project, and the log density in a thousand lines of code (KLOC). 
All of the selected projects use Apache Log4j library~\cite{log4x} as the logging statements' wrapper function, which includes six log verbosity levels:\textit{ fatal, error, warn, info, debug,} and \textit{trace}. 
We observed that although all these projects are from the Apache umbrella, they are from different domains, are developed by different teams, and incorporate different logging practices.

\subsection{RQ3.I: LACCP Evaluation}\label{logprediction}
\subsubsection{Evaluation Metrics.}
For evaluating the performance of the logging statement prediction, we utilize Precision, Recall, F-Measure, and Balanced Accuracy. 
\begin{itemize}
\item \textit{Precision} is the ratio of the correctly predicted log statements ($t_p$) to the total number of predictions ($t_p+f_p$), $Precision= \frac{t_p}{t_p+f_p}$.

\item \textit{Recall} is the ratio of $t_p$ over the total number of log statement instances detected and not-detected ($t_p+f_n$), $Recall= \frac{t_p}{t_p+f_n}$.

\item In order to confirm the balance between the \textit{Precision} and \textit{Recall} values, we also calculate \textit{F-Measure} which is the harmonic average of the \textit{Precision} and \textit{Recall}, $F-Measure= 2 \times \Big(\frac{Recall \times Precision}{Recall + Precision}\Big)$. 

\item In addition, to ensure our results are impartial towards the imbalanced datasets~\cite{zhu2015learning}, we also measure \textit{Balanced Accuracy (BA)}, which is the average of logged and unlogged methods that are correctly predicted, $BA= \frac{1}{2} \times (\frac{t_p}{t_p+f_n} + \frac{t_n}{t_n+f_p})$. 
\end{itemize}

For method-level clone detection in our studied systems, we use the same pre-trained learning model provided in~\cite{saini2018oreo} as our baseline. 
To ensure variability in the dataset, the machine learning engine of the clone detector is trained on 50K randomly-selected Java projects from Github~\cite{saini2018oreo}. 
Randomly, 80\% of each project is selected for training and the remaining 20\% for testing. 
Additionally, one million pairs from the training set of 50K projects are kept separately for validation to avoid bias and overfitting~\cite{caruana2001overfitting}. 
To ensure a fair comparison, all of the approaches use the same trained model for clone detection, and theretofore, the advantage of LACCP becomes visible from its log-aware source code feature selection and calculation.  

\subsubsection{Experiments and Results.}
Our goal in this experiment is to show LACCP's performance in predicting the log locations for clone pairs compared to LACC and Oreo. 
For experiment design, we first extract all the methods with logging statements and then remove their logging statements.
Next, we check which pairs are detected as pairs after their logging statemented are removed, \textit{i.e.}, $(MD_{i'},MD_{j'})$, which serves as actual true positive cases, $t_p$, and which method are not detected as clone pairs, \textit{i.e.}, $t_n$. 
This process ensures that the methods are detected as clones regardless of their logging statements, which are considered as \textit{ground truth} for our comparison. 
In this case that there are no logging statements in the methods, all of the approaches (Oreo, LACC, and LACCP) will have similar performance in clone detection as log-aware feature calculation will have an effect on the feature values only if logging statements are present in the code snippet. 
Later, we run Oreo, LACC, and LACCP on the collected pairs as $(MD_{i},MD_{j'})$ and evaluate their performance compared to the \textit{ground truth}. 
The experimented scenario resembles the situation when the developer is composing a new source code snippet without a logging statement ($MD_{j'}$), and we look in the code base to find clones with logging statements ($MD_{i}$) and provide logging suggestions to the developer. 

Table~\ref{laccp_result} shows \textit{Precision}, \textit{Recall},  \textit{F-Measure}, and \textit{BA} for the three approaches. 
For log statement prediction for seven different projects, the measured values for \textit{BA} are in the range of $(78\%,92\%)$ for Oreo, $(81\%,97\%)$ for LACC, and $(95\%,100\%)$ for LACCP. 
For example, for Hadoop project, with the number of \textit{ground truth} positive cases of 879, and \textit{ground truth} negative cases of 235, after removing logging statements for $MD_j$s, Oreo had 16 false positives ($f_p$), a significant number (319) of false negatives ($f_n$), and 560 true positives ($t_p$), achieving `imbalanced' Precision and Recall of 97.22\% and 63.71\%, respectively. 
In comparison, for LACCP: ($f_p=18$), ($f_n=17$), and ($t_p=862$), yielding Precision and Recall of 97.95\% and 98.07\%. 
Although LACC improves on Oreo by partially reducing the number of false negatives, it still suffers from a high rate of false negatives, which hurts its Recall scores when compared to LACCP.  
Overall, considering the average of BA values, LACCP brings 15.60\% and 5.93\% improvement over Oreo and LACC, respectively, across the experimented projects. 
The higher accuracy that LACCP brings enables us to provide more accurate clone-based log statement suggestions.

\afterpage{
\begin{landscape}
%\begin{sidewaystable}
\begin{table*}
  %\hspace*{-5mm}
 \vspace*{3.5cm}
  \scriptsize
\begin{tabular}{p{1.1cm}|p{.2cm}p{.2cm}p{.2cm}p{.2cm}p{.4cm}p{.4cm}p{.4cm}p{.4cm}|p{.2cm}p{.2cm}p{.2cm}p{.2cm}p{.4cm}p{.4cm}p{.4cm}p{.4cm}|p{.2cm}p{.2cm}p{.2cm}p{.2cm}p{.4cm}p{.4cm}p{.4cm}p{.4cm}| p{.7cm}p{.80cm}}
    \midrule \rowcolor{blue!10}
   {\bfseries Proj.\textbackslash Tool} &
    \multicolumn{8}{c}{\bfseries Oreo~\cite{saini2018oreo}  (X)}& 
    \multicolumn{8}{c}{\bfseries LACC~\cite{gholamian2020logging}  (Y)}&
    \multicolumn{8}{c}{\bfseries LACCP  (Z)}&
    \multicolumn{2}{c}{\bfseries Improvement \%}
        \\ %\cline{3-4}
     \rowcolor{blue!10}  & $tp$ & $tn$& $fp$ & $fn$ & $P$ & $R$& $F$& $BA$& $tp$ & $tn$& $fp$ & $fn$ & $P$ & $R$& $F$& $BA$& $tp$ & $tn$& $fp$ & $fn$ & $P$ & $R$& $F$& $BA$& Z{\tiny(over)}X & Z{\tiny(over)}Y\\

    \midrule
    
       HD& 560 & 219& 16 & 319 &97.22 &63.71& 78.98& 78.45& 825&218&17&54&97.98 &93.86&95.87&93.31 &862& 217& 18& 17 & 97.95& 98.07 & 98.01 & 95.20 & 21.35  & 2.03 \\   
\midrule
       ZK& 47 & 33& 0 &17& 100 & 73.44 &84.68& 86.72&  62&27&6&2& 91.18&96.86&93.94 &89.35& 61& 32& 1 & 3& 98.39 & 95.31 & 96.83 & 96.14& 10.86&7.60\\    
       \midrule
CS& 430 & 222& 22 & 184 & 95.13 &70.03& 80.68& 80.51& 441&223&21&173&95.45 &71.82&81.97&81.61 & 610& 239& 5& 4 & 99.19& 99.35 & 99.27 & 98.65 & 22.53& 20.88\\
          \midrule    
HB& 377 & 182& 13 & 115 & 96.67&76.62& 85.49& 84.98& 479&190&5&13& 98.97&97.36&98.16&97.04 &492& 195& 0& 0 & 100& 100 & 100 & 100 & 17.67& 3.05\\ 
                            
                 \midrule    
HI &586  & 99& 6 & 66 &98.99 &89.88& 94.21& 92.08& 643&100&5&9&99.22 &98.62&98.92&96.93 &648& 100& 5& 4 & 99.23 &99.39  & 99.31 & 97.31 & 5.68 &	0.39\\ 
                             \midrule                
CL& 633 & 270& 17 & 164 &97.38 &79.42& 87.49& 86.75& 746&285&2&51&99.73 &99.73&96.57&96.45 &797& 285& 2& 0 & 99.75&100 & 99.87 & 99.65 &14.87&3.32\\ 
                             \midrule                                            
                                                                                                                MQ& 338 & 207& 4 & 131 &98.83 &72.07& 83.35& 85.09& 423&210&1&46 &99.76&90.19&94.74& 94.86 &463& 209& 2& 6 & 99.57& 98.72 & 99.14 & 98.89 & 16.22&	4.25\\ 
\midrule
\textbf{Total/Avg}& 2971&1232&78&996&97.75&75.02&84.98&84.94& 3619&	1253&	57&	348&	97.47&	92.63&	94.31&	92.79& 3933&	1277&	33&	34&	99.15&	98.69&	98.92& 97.98 & 15.60&5.93\\                                     
       
    %------
  \bottomrule
  \end{tabular}
  \caption{The table shows the value of $t_p$, $t_n$, $f_p$, and $f_n$ for the three approaches. We also show Precision (P), Recall (R), F-Measure (F), and BA for the three methods of log prediction. The general trend on how the methods perform is observable on F-Measure, and BA metrics, as the values increase, \textit{i.e.}, Oreo < LACC < LACCP.}
  \label{laccp_result}
%\vspace*{-4mm}
\vspace*{2mm}
\end{table*}
%\end{sidewaystable}
\end{landscape}
}

\subsubsection{Qualitative Comparison.} Intuitively, because LACCP performs log-aware feature selection and calculation, it understands the log feature changes while processing the source code, and it achieves a higher log-aware clone prediction performance compared to the general-purpose clone detector, Oreo~\cite{saini2018oreo}. 
In addition, LACCP, in contrast to LACC that only addresses \textbf{SI} and still suffers from a significant number of mis- and not-detected clones, seeks to simultaneously address both \textbf{SI} and \textbf{SII} and reaches balanced Precision and Recall scores, and therefore, higher F-Measure and Balance Accuracy (BA) values compared to LACC. 

Orthogonal to our research, prior efforts, such as \cite{zhu2015learning}, \cite{jia2018smartlog}, and~\cite{li2020shall}, have proposed learning approaches for logging statements' \textit{location} prediction, \textit{i.e.}, \textit{where to log}. 
The approaches in \cite{zhu2015learning}, \cite{jia2018smartlog} are focused on error logging statements (ELS), \textit{e.g.}, log statements in \textit{catch clauses}, and are implemented and evaluated on C\# projects. 
Li \textit{et al.}~\cite{li2020shall} provide log location suggestions by classifying the logged locations into six code-block categories. Different from these prior works, our approach does not distinguish between error and normal logging statements, is evaluated on open-source Java projects, and leverages logging statement suggestions at method-level by observing logging patterns in similar code snippets, \textit{i.e.}, clone pairs. 
In addition, none of the aforementioned studies have tackled the automation of the log statements descriptions, which we have also proposed in this research, and will evaluate in the following.

\vspace{1mm}
\begin{tcolorbox}[breakable, enhanced,after=\ignorespacesafterend\par\noindent]
\textbf{Finding} \textbf{1.} \textit{Augmenting general-purpose clone detection approaches with log-aware features is necessary and beneficial for log statement automation, as we aim to predict a logging statement for a code snippet that is unlogged, by looking for its similar clones, which are logged.\looseness=-1}
\end{tcolorbox}

\begin{tcolorbox}[breakable, enhanced,after=\ignorespacesafterend\par\noindent]
\textbf{Finding} \textbf{2.} \textit{Our results show that LACCP outperforms Oreo and LACC in Balanced Accuracy values by 15.60\% and 5.93\%, respectively.}
\end{tcolorbox}

\subsection{RQ3.II: LSD Evaluation}\label{eval}
In this section, we measure the performance of (NLP CC'd) for logging statements' \textit{description} prediction. 
If we can achieve a satisfactory performance, an automated log description predictor that can suggest the description of logging statements will be of great help, as it can significantly expedite the software development process and improve the quality of logging descriptions. 
To measure the accuracy of our method in suggesting the log description, we utilize BLEU~\cite{papineni2002bleu} and ROUGE~\cite{lin2004rouge} scores. 
The rationale behind using these scores is that they are well-established for validating the usefulness of an auto-generated text~\cite{bahdanau2014neural,rush2015neural,lin2004rouge,zhou2018neural,graham2015re}. 
In particular, prior software engineering and machine learning research have used these scores for tasks such as comment and code suggestion~\cite{allamanis2018survey} and description prediction~\cite{he2018characterizing}. 
In addition, they are intuitively equivalent to \textit{Precision} and \textit{Recall} for evaluating auto-generated text.

\subsubsection{BLEU Score}
BLEU, or the \textit{Bilingual Evaluation Understudy}, is a score for comparing a candidate text to one or more reference texts. 
BLEU score is used to evaluate text generated for a series of natural language processing tasks~\cite{bahdanau2014neural,luong2015effective}. 
BLEU score can measure the similarity between a candidate and a reference sentence. 
In our experiments, we regard the logging description generated by finding the code clone pair snippet and then predicted by the NLP model as the candidate description, while we refer to the original logging description written by the developer as the reference description. 
BLEU measures how many n-grams (\textit{i.e.}, tokens) in the candidate logging description appear in the reference, which makes it comparable to \textit{`Precision'}.
BLEU is evaluated as:
%\vspace*{-1mm}
\begin{equation}\label{bleu_f}
BLEU= BP \times exp\bigg( \sum_{n=1}^{N} w_n \times \log{p_n} \bigg)
\vspace*{-1mm}
\end{equation}
where BP is a \textit{`brevity penalty'} that penalizes if the length of a candidate (in number of tokens) is shorter than the reference:\vspace{-2mm}
%Specifically, $BP$ is calculated as follows:\vspace{-2mm}
\[\scriptstyle        
BP = 
     \begin{cases}
       \text{1} &\quad\text{if c $>$ r }\\
       \displaystyle\text{$e^{1- \frac{r}{c}}$} &\quad\text{if c $\leq$ r }\\
     \end{cases}
\]
where $r$ is the length of the reference, and $c$ is the length of the candidate. 
In Formula~\ref{bleu_f}, N is the maximum number of n-grams used in the experiment; $p_n$ is the modified n-gram Precision, which is the ratio of the number of tokens from the candidate logging description which occur in the reference description to the total number of token in the candidate; and $w_n$ is the weight of each $p_n$. 
For example, BLEU-1 means the BLEU score considering only the 1-grams in the calculation, where $w_1$ = 1 and $w_2$ = $w_3$ = $w_4 = ...$ = 0.\looseness=-1

From the definition of BLEU, we know that the higher the BLEU, the better the logging statement description prediction performance. 
The range of BLEU is [0, 1], or as a percentage value (\textit{i.e.}, [0, 100]). 
Thus, if the candidate logging description does not contain any of the reference's n-grams, then the BLEU score is 0. 
On the contrary, if all of the candidate tokens appear in the reference, the BLEU score is 100. 
An additional enhancement to the BLEU score is to calculate cumulative scores as it gives a better sense of the sentence level structure similarity between the candidate and reference descriptions.  
Cumulative BLEU scores refer to the calculation of individual n-gram scores at all orders from 1 to N and weighting them by calculating the weighted geometric mean. 
The cumulative and individual 1-gram BLEU use the same weights, \textit{e.g.}, (1, 0, 0, 0). 
The 2-gram weights assign a 50\% to each of 1-gram and 2-gram (\textit{e.g.}, (0.5, 0.5, 0, 0)), and the 3-gram weights are 33\% for each of the 1, 2 and 3-gram scores (0.33, 0.33, 0.33, 0). 

We provide two sets of BLEU scores: 1) the BLEU scores generated by simply borrowing the LSD from the clone pair as it is (\textit{i.e.}, the LSD of $MD_i$ as a candidate, and the LSD of $MD_j$ as reference), \textbf{No-NLP}, and 2) the BLEU scores for the LSD predicted by the NLP CC'd model by considering only one previous token of the input LSD string, LSD($MD_i$), for predicting the next token, \textit{i.e.}, \textbf{NLP-1}. 
The NLP-1 is the output of the LSTM model, and the original LSD of $MD_j$ is considered as the reference for score evaluation.
The \textbf{No-NLP (X)} and \textbf{NLP-1 (Y)} columns in Table~\ref{fig:total_compare} outline the cumulative BLEU scores for No-NLP and NLP-1. 
The BLEU-1 scores for all of the evaluated projects are higher than 47\%. 
The highest BLEU-1 score belongs to Hive, 92.50\%, which means that 92.50\% of the tokens in the generated logging description of the candidate can be found in the ground truth, \textit{i.e.}, the logging description placed in by the developer. 
This observation implies that in most cases for this project (Hive), developers have reused the logging descriptions from the existing log statements with minor modifications. 
The NLP-1 model improves the performance of the LSD prediction across all projects. 

\subsubsection{ROUGE Score}%\vspace*{-1mm}
ROUGE~\cite{lin2004rouge} stands for \textit{Recall-Oriented Understudy for Gisting Evaluation}. 
It includes measures to automatically determine the quality of an auto-generated description by comparing it to other (ideal) descriptions created by humans, \textit{i.e.}, developers in our case.
Formally, ROUGE-N is an n-gram Recall between a candidate description and a reference description. 
ROUGE-N is computed as follows:
\begin{align*}\vspace*{-2mm}
\scriptstyle ROUGE-N& = \frac{\sum\limits_{gram_N \in Ref} Count_{match}(gram_N)}{ \sum\limits_{gram_N \in Ref} Count(gram_N) }\vspace*{-3mm}
\end{align*}
where $N$ represents the number of overlapping grams that have to match in reference and candidate descriptions; 
%$n$ represents the length of the n-gram, $gram_n$; 
$Ref$ is the reference description; $count_{match}$ $(gram_N )$ is the maximum number of n-grams co-occurring in the candidate and the reference; and $count (gram_N)$ is the number of n-grams in the reference.
ROUGE is similar to \textit{`Recall'}, which measures how many n-grams in the reference appear in the candidate logging statement. 
For example, ROUGE-2 explains the overlap of 2-grams between the candidate and reference descriptions. 
%ROUGE-L is a longest Common Subsequence (LCS) based statistics. 
ROUGE-L is a statistic calculated based on the Longest Common Subsequence (LCS). 
ROUGE-L takes into account the sentence level structure and similarity, and thus, identifies the longest co-occurring sequence of continuous n-grams. 
Analogous to BLEU, the range of ROUGE is [0, 1], with 1 being the perfect score, \textit{i.e.}, the candidate description contains all of the reference's n-grams. 
Similar to BLEU, we provide two sets of ROUGE scores: 1) \textbf{No-NLP}, and 2) \textbf{NLP-1}. 
Table~\ref{fig:total_compare} compares ROUGE scores for No-NLP versus NLP-1. 
Similar to BLEU, the NLP model suggests LSDs with higher ROUGE scores in all cases when compared to the LSD obtained from the clone pair.

\subsubsection{NLP Prediction Example}  
%s\textbf{NLP prediction example.} 
To illustrate how the NLP model can improve BLEU and ROUGE scores, we provide the following real example. 
In the \textbf{training set} for CloudStack we have the following LSDs: \textit{``\textbf{successfully deleted} condition''} (1x), \textit{``elastistor volume \textbf{successfully deleted}''} (3x), and \textit{`\textbf{`successfully created} floating ip''} (1x). 
During \textbf{testing}, the retrieved LSD from the clone pair, $MD_i$, is \textit{``successfully created floating ip''}, and the original LSD that we are aiming to predict, \textit{i.e.}, the reference LSD of $MD_j$, is \textit{``\textbf{successfully deleted} floating ip''}. 
Because in the training set the token \textit{``deleted''} appears four times (\textit{i.e.}, 1x+3x=4x) right after the token \textit{``successfully''}, whereas \textit{``created''} appears only once (1x) immediately after \textit{``successfully''}, the NLP model assigns a higher probability for observing \textit{``deleted''} after \textit{``successfully''}. 
Therefore, the NLP model allows us to see highly probable next tokens that appear in the training set as a whole that might not necessarily happen in the LSD retrieved from the clone pair.

\subsubsection{LSD Sequence Prediction - NLP-1 vs. NLP-3} 
Because LSDs are a sequence of natural language tokens~\cite{he2018characterizing,gholamian2021naturalness}, we hypothesize that considering additional prior tokens for predicting the next token will achieve higher performance. 
As such, in NLP-3, in contrast to NLP-1 that we consider only the one prior token, we take into account the sequence of three prior tokens in predicting the next token. 
Additionally, because the outputs of the NLP CC'd model are \textit{softmax} probabilities (Figure~\ref{lsd_nlp}), we report the top-3 probabilities for recommending the next token and then confirm which one achieves higher BLEU and ROUGE scores. 
This approach resembles the scenario in which a list of high probable next tokens is suggested to the developer while composing the LSD. 
In Table~\ref{fig:total_compare}, \textbf{NLP-3 (Z)} illustrates the improvement that NLP-3 brings compared to NLP-1. 
The rationale behind choosing three prior tokens and not considering longer sequences for LSD prediction is that LSDs are naturally shorter than English text sentences and considering more than four continuous tokens, \textit{i.e.}, three prior tokens and the one token under prediction, would result in a minimal gain or even might cause inaccuracies for shorter than four-token LSDs.  
In sum, our LSTM model generates candidate LSDs with higher BLEU and ROUGE scores in all cases when compared to the LSD borrowed from the clone pair, and additionally, we enhance the NLP CC'd performance by leveraging a sequence of tokens for LSD prediction.  
Our experiments signify the benefits of the collaboration of LACCP and NLP CC'd for LSD prediction.

\afterpage{
%\diagbox[innerwidth=\textwidth*2/17]{Projects}{ \qquad Scores (\%)}
\begin{landscape}
\begin{table*}
\vspace*{1cm}
% \hspace*{-5mm}
  \scriptsize
  %\footnotesize
\begin{tabular}{p{.5cm}|p{.4cm}p{.4cm}p{.4cm}p{.4cm}p{.4cm}p{.4cm}p{.4cm}p{.4cm}|p{.4cm}p{.4cm}p{.4cm}p{.4cm}p{.4cm}p{.4cm}p{.4cm}p{.4cm}|p{.4cm}p{.4cm}p{.4cm}p{.4cm}|P{.5cm}P{1cm}}
    \toprule \rowcolor{blue!10}
    {\bfseries Prj.} &
    \multicolumn{8}{c}{\bfseries No-NLP \% (X)}&
    \multicolumn{8}{c}{\bfseries NLP-1 \% (Y)}&
    \multicolumn{4}{c}{\bfseries NLP-3 \% (Z)}&
    \multicolumn{2}{c}{\bfseries Improv. Z{\tiny(over)}X}
        \\ %\cline{3-4} 
 \rowcolor{blue!10}
    & B-1 & B-2& B-3 & B-4 & R-1 & R-2& R-3 & R-L& B-1 & B-2& B-3 & B-4 &R-1 & R-2& R-3 & R-L& B-1 & B-4& R-1 & R-L& B-4 & R-L \\
       
    \midrule
    %------
    HD  & 58.42& 48.09&39.43&32.91&59.22&36.00&20.13&59.07&    
    58.87&48.63&40.18&33.27 & 53.77&32.89&59.40&58.92& 60.22&34.97&
    62.39&62.18& 6.26&5.26 \\     \midrule
    ZK & 64.15 &  56.85&47.02&41.59&63.45&46.31&25.64&63.45&
    65.15 & 58.26&48.53&43.35&66.41&50.00&30.44&66.41& 66.05&43.46
    &68.53& 68.53& 4.50&8.01 \\     \midrule
   CS & 47.45& 39.50&33.98&29.42&49.16&33.30&22.23&
   48.68&47.87&39.89&34.38&29.79
   &50.45&34.50&23.19&49.97& 49.92& 32.26& 53.53&53.02&9.65&8.92 \\    \midrule
      HB & 82.67& 77.37&71.78&56.80&83.15&71.68&60.86&83.06&83.18&78.02&72.40&47.38
   &84.32&73.37&62.15&84.23& 83.67& 58.38& 84.64&84.53&2.78&1.77 \\    \midrule
         HI & 92.50 & 91.07&89.62&70.57&92.61&88.21&85.36&92.57&
         92.55&91.10&89.64&70.58&92.65&88.30&85.41&92.61& 92.64&
70.61&92.73&92.69&0.06&0.13 \\     \midrule
CL&66.36&55.65&49.15&41.17&65.34&42.48&25.24&64.65&66.51&55.71&49.20&41.20
   &65.56&42.68&25.50&64.87&67.40&43.12&67.31&66.59 &4.74&3.00\\     \midrule
   MQ&69.94&60.14&47.54&38.86&68.43&43.56&22.58&68.27&70.51&60.40&47.74&39.04
   &69.55&44.00&22.58&69.39&71.15& 40.18&71.25&71.08&3.40&4.12\\

  \midrule
    \textbf{Avg.} & 68.78&	61.24&54.07&44.47&68.77&51.65&37.43&	68.54& 68.39&59.25&57.23&47.08&68.96&52.25&44.10&69.49&70.15&46.14&	71.48&71.23&4.48&4.46\\
  \hline
  \end{tabular}
   \caption{BLEU (B) and ROUGE (R) scores for No-NLP, NLP-1, and NLP-3 are included side-by-side for each project. The NLP model improves the scores across the board. For example, for MQ, the No-NLP B-1 and R-1 scores are 69.94 and 68.43, respectively, and the values increase to 70.51 and 69.55 for the NLP-1 model, and furthermore, rise to 71.15 and 71.25 for the NLP-3 model.}
    \label{fig:total_compare}

\end{table*}
%}    

\begin{table*}[h]
  \vspace*{3mm}

  \scriptsize
\begin{tabular}{p{.9cm}|p{.4cm}p{.4cm}p{.4cm}p{.4cm}|p{.4cm}p{.4cm}p{.4cm}p{.4cm}|p{.4cm}p{.4cm}p{.4cm}p{.4cm}|p{.4cm}p{.4cm}p{.4cm}p{.4cm}|p{.6cm}p{.9cm}}
    \toprule \rowcolor{blue!10}
    %\multirow{2}{*}
    {\bfseries Projects} &
    \multicolumn{4}{c}{\bfseries He \textit{et al.}~\cite{he2018characterizing} \% (W)}& 
    \multicolumn{4}{c}{\bfseries No-NLP \% (X)}&
    \multicolumn{4}{c}{\bfseries NLP-1 \% (Y)}&
     \multicolumn{4}{c}{\bfseries NLP-3 \% (Z)}&
    \multicolumn{2}{c}{\bfseries Improvement \%}
        \\ %\cline{3-4}  
        \rowcolor{blue!10}
    & B-1 & B-4& R-1 & R-L & B-1 & B-4& R-1 & R-L& B-1 & B-4& R-1 & R-L &B-1 & B-4& R-1 & R-L& X{\tiny(over)}W & Z{\tiny(over)}X\\
       
    \hline
    %------
    HD  & 36.59& 16.96&36.88&	36.24	& 51.74 & 30.21 & 56.97 & 56.49 &    52.95&31.27&58.72&58.22 & 53.77&32.89&59.40&58.92& 54.27 & 4.90 \\
   CS & 47.60 & 27.57 &47.11	&46.05	& 50.20&33.11& 55.81& 54.78 &50.99&33.95&57.60&56.68 &53.89&37.21&61.07&59.98&  15.19 & 9.41\\
   HB & 37.69 & 18.28 &	38.47 &	37.71 &49.70&30.06& 56.98&56.59& 50.31 &30.75 & 58.50& 58.08& 52.32&33.67&60.00&59.61&  46.30 & 6.35\\
    HI & 40.78 & 23.04 &	40.58 &	40.08 &51.99&36.13& 54.95&53.36& 52.47 &36.31 & 55.43& 53.84& 54.16&38.74&58.70&56.85&34.31&6.12\\
 CL & 51.98  &30.74 &50.23	 &	49.62 &60.03&34.77&63.86&63.02&60.56 &35.01 & 63.97& 63.13& 62.54&38.44&66.79&65.91&21.42&5.41\\  
  \midrule
    \textbf{Average} &43.06&23.18&42.85&42.11&
    52.73&32.86&57.71&56.84&
    53.46&33.46&58.84&57.99&
    55.34&36.19&61.19&60.25&32.38&6.41 \\
  \bottomrule
  \end{tabular}
   \caption{BLEU (B) and ROUGE (R) scores comparison for~\cite{he2018characterizing}, the LSD from code clone with no modification, \textit{i.e.}, No-NLP, considering only one prior token in prediction, NLP-1, and considering a sequence of three prior tokens, NLP-3. The `Improvement' column shows the percentage that No-NLP improves on prior work, and how much NLP-3 improves over No-NLP. On average, NLP-3 makes 40.86\% improvement over~\cite{he2018characterizing} (Z{\tiny(over)}W).}
    \label{table_he_compare}
\end{table*}
\end{landscape}
}

\subsubsection{Results Review and Comparison}
In Table~\ref{fig:total_compare}, the BLEU and ROUGE scores gradually decrease as the n-grams grow longer. 
For example, for \textbf{NLP-1}, BLEU-1 for Hadoop is 58.87, while the corresponding BLEU-4 is 33.27. 
This observation is expected because the BLEU-4 score considers the match of 4 consecutive tokens (\textit{i.e.}, 4-grams) versus BLEU-1, which only considers matching 1-grams.\looseness=-1

To provide an intuitive understanding of how good our BLEU and ROUGE scores are, we compare as follows: the BLEU-4 scores of our NLP method outperform prior efforts in~\cite{loyola2017neural} and~\cite{he2018characterizing}.
From a practitioner perspective, the satisfactory BLEU-4 scores reported in the state-of-the-art code summarization paper~\cite{loyola2017neural}, ranges from 6.4\% to 34.3\%, which are lower than our reported values. 
The authors in~\cite{loyola2017neural} showed that with their achieved BLEU scores, their auto-generated code summaries are both \textit{fluent} and \textit{informative} for the human reader. 

For a direct comparison, we use He \textit{et al.}'s sample data available in~\cite{urlhecompare}. 
The scores only exist for five projects of our interest (Hadoop, CloudStack, HBase, Hive, and Camel)~\cite{he2018characterizing}, and their approach includes pairs of \textit{`(code, log)'}, with \textit{`code'} indicating the ten lines of code preceding the logging statement \textit{`log'}.
As such, to employ our approach, we perform initial preprocessing to wrap the \textit{`code'} segment inside a dummy method, and then we utilize LACCP and NLP CC'd for LSD suggestions, and then compare them with the provided \textit{`log'}. 
Table~\ref{table_he_compare} summarizes the scores for~\cite{he2018characterizing} and our approach. 
He \textit{et al.}~\cite{he2018characterizing} achieve 36.59 and 37.69 BLEU-1 scores for Hadoop and HBase, respectively. 
In comparison, our No-NLP approach achieves 51.74 and 49.70 for BLEU-1 scores for Hadoop and HBase, respectively. 
Similarly, our ROUGE scores outperform prior work. 
We contribute this higher performance to the more sophisticated search of clone pairs compared to employing the ten preceding lines of the code utilized in~\cite{he2018characterizing}. 
Table~\ref{table_he_compare} also provides the NLP-1 and NLP-3 scores, which further improve the No-NLP ones. 
The NLP model is successful in remembering the general LSD patterns in each project and further enhances the LSD suggestions. 
Another observation we made is that NLP CC'd values are generally lower on the sample data~\cite{urlhecompare} in Table~\ref{table_he_compare} than the values in Table~\ref{fig:total_compare}, as we hypothesize method-level clone detection provides a better context for LSD prediction than selecting the ten preceding lines of code in the sample data~\cite{he2018characterizing}.

\vspace{1mm}
\begin{tcolorbox}[breakable, enhanced]
 \textbf{Finding} \textbf{3.} \textit{The additional context provided through finding similar code snippets can be borrowed as a starting point for LSD automation and further augmented with deep-learning NLP approaches.}
\end{tcolorbox}

\begin{tcolorbox}[breakable, enhanced]
 \textbf{Finding} \textbf{4.} \textit{Our LSD prediction approach, on average, achieves 32.38\% improvement over the prior work ($\frac{X}{W}$) and 6.41\% improvement over the No-NLP version ($\frac{Z}{X}$).\looseness=-1}
\end{tcolorbox}

\section{Case Study}\label{casestudy}
Yuan \textit{et al.}~\cite{yuan2012characterizing} showed that developers spend a significant amount of time revising logging statements for system dependability tasks, such as postmortem failure analysis. 
In this case study for the Hadoop project (Listing~\ref{w_o_log_lvl_2}), we investigate the code snippets that have logging statements updates or revisions after a problem was detected in the software systems. 
Then, we try to find code clones based on that snippet of the source code and see if we could have predicted the logging statement description prior to the failure that could have saved engineering time and trial-and-error cycle. 
We use \textit{`git blame'} to assign commit number and data and time of the commit, and we look for clones in the portion of the code which was developed and checked in prior to the log statement fixes (\textit{i.e.}, for clone detection we rely on the code which was previously developed while the new code is being composed). 
We found a JIRA ticket for the YARN subsystem of Hadoop, YARN-985~\cite{jirahadoop}. 
Code Snippets 1 and 2 are clone pairs in Listing~\ref{w_o_log_lvl_2}. 
Snippet 2 went through two logging updates in two different \textit{git commits} in the Year 2014, highlighted in orange and red, respectively. 
These modifications could have been avoided if the engineer developing the logging statement to Code Snippet 2 had access to observe the logging description from its clone pair, Code Snippet 1 and NLP CC'd predictions for LSD suggestions.

\begin{figure}[h]
%\vspace*{-4mm}
\begin{minipage}{\linewidth}
%\belowcaptionskip=-10pt
\begin{lstlisting}[linebackgroundcolor={%
    \ifnum\value{lstnumber}=1
            \color{blue!10}
    \fi
    \ifnum\value{lstnumber}=11
            \color{blue!10}
    \fi    
    },
caption={Case study from JIRA; two log changes.},label={w_o_log_lvl_2},style=base]
//Snippet 1, FairScheduler.java.
2013-12-13    private synchronized void removeApplicationAttempt(
2012-07-13        ApplicationAttemptId applicationAttemptId,
2014-01-10        RMAppAttemptState rmAppAttemptFinalState, boolean keepContainers) {
!!2012-07-13!!          LOG.info("Application " + applicationAttemptId + " is done." +
!!2012-07-13!!          " finalState=" + rmAppAttemptFinalState);
2014-08-12            SchedulerApplication<FSAppAttempt> application =
2014-01-10            applications.get(applicationAttemptId.getApplicationId());
2014-08-12            FSAppAttempt attempt = getSchedulerApp(applicationAttemptId);
                      ...
//Snippet 2, CapacityScheduler.java.
2013-12-13    private synchronized void doneApplicationAttempt(
2011-08-18        ApplicationAttemptId applicationAttemptId,
2014-01-10        RMAppAttemptState rmAppAttemptFinalState, boolean keepContainers) {
^2014-01-02^     LOG.info("Application Attempt " + applicationAttemptId + " is done." +
@2014-09-12@      " finalState=" + rmAppAttemptFinalState);
2014-01-10       FiCaSchedulerApp attempt = getApplicationAttempt(applicationAttemptId);
2014-05-22       SchedulerApplication<FiCaSchedulerApp> application =
2014-01-10       applications.get(applicationAttemptId.getApplicationId());
                      ...
\end{lstlisting}
\end{minipage}
%\vspace*{-7mm}
\vspace{2mm}
\end{figure}

\section{Discussion}\label{discussion}
\subsection{Log Verbosity Level (LVL) and Variables (VAR)}
%\textbf{Log verbosity level (LVL) and variables (VAR).} 
Our approach is reasonably extendable to predict LVL and VAR alongside the LSD suggestion. 
Since we have access to the source code of the method that we are predicting the logging statement for and its clone pair, a reasonable starting point is to suggest the same LVL as of its clone pair, and then augment it with additional learning approaches such as~\cite{li2017log,anu2019approach,li2021deeplv} for more sophisticated LVL prediction. For example, our analysis for the evaluated projects in Figure~\ref{fig_lvl} shows that code clones match in their verbosity levels in the range of (92, 97)\%. 
In addition, we also hypothesize that some of the LVL mismatches in the clone pair LVLs are due to the log-related issues~\cite{hassani2018studying} that our clone-based approach has uncovered.  
For VAR prediction, our approach can be augmented with deep learning~\cite{liu2019variables} and static analysis of the code snippet under consideration~\cite{yuan2012improving} to include log variables suggestions alongside the predicted LSD. 
\afterpage{
\begin{figure}
\vspace{2mm}
        \begin{subfigure}[b]{0.23\textwidth}
            \includegraphics[width=1\textwidth]{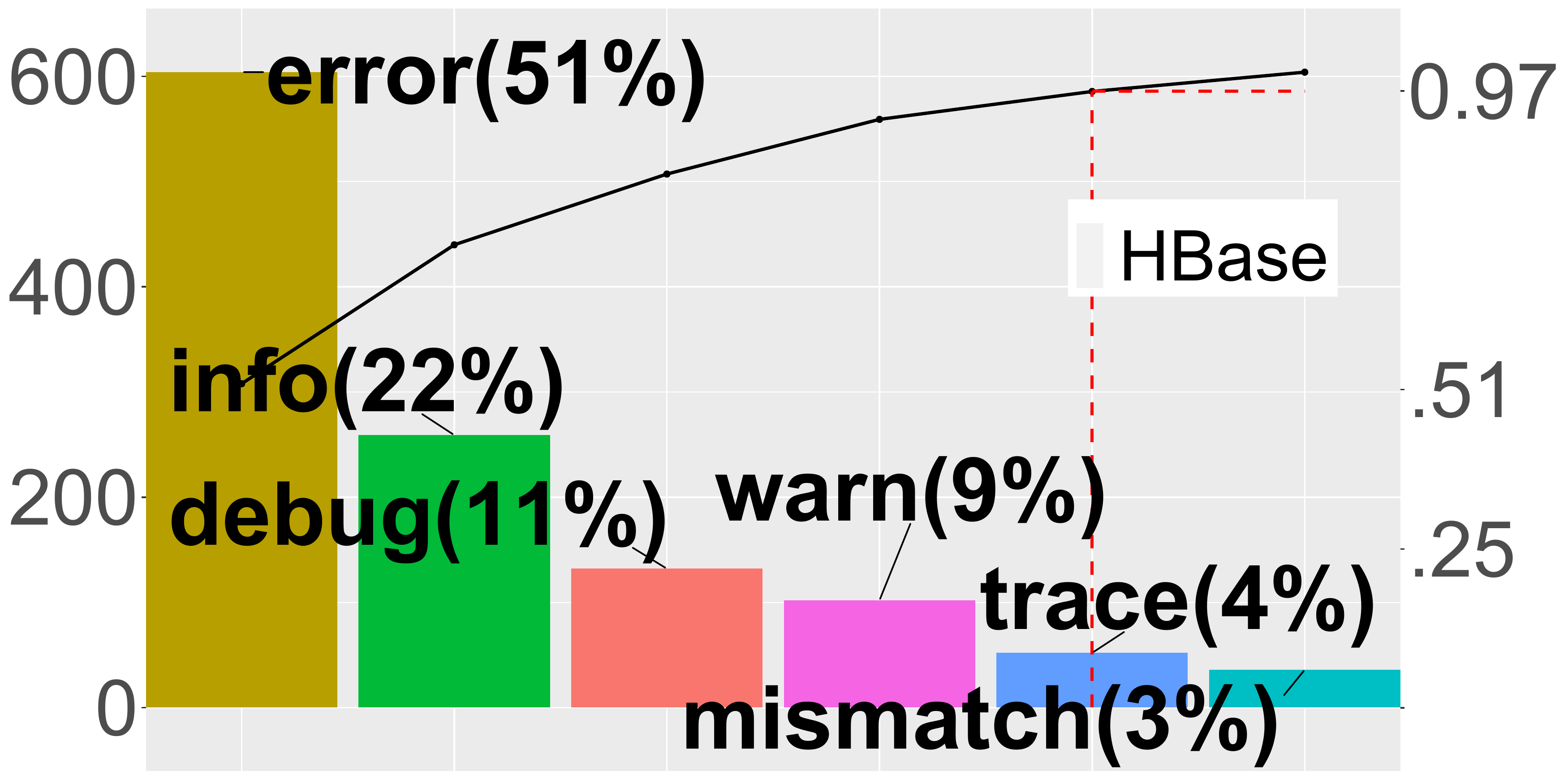}
        \end{subfigure}  
               \begin{subfigure}[b]{0.23\textwidth}
            \includegraphics[width=1\textwidth]{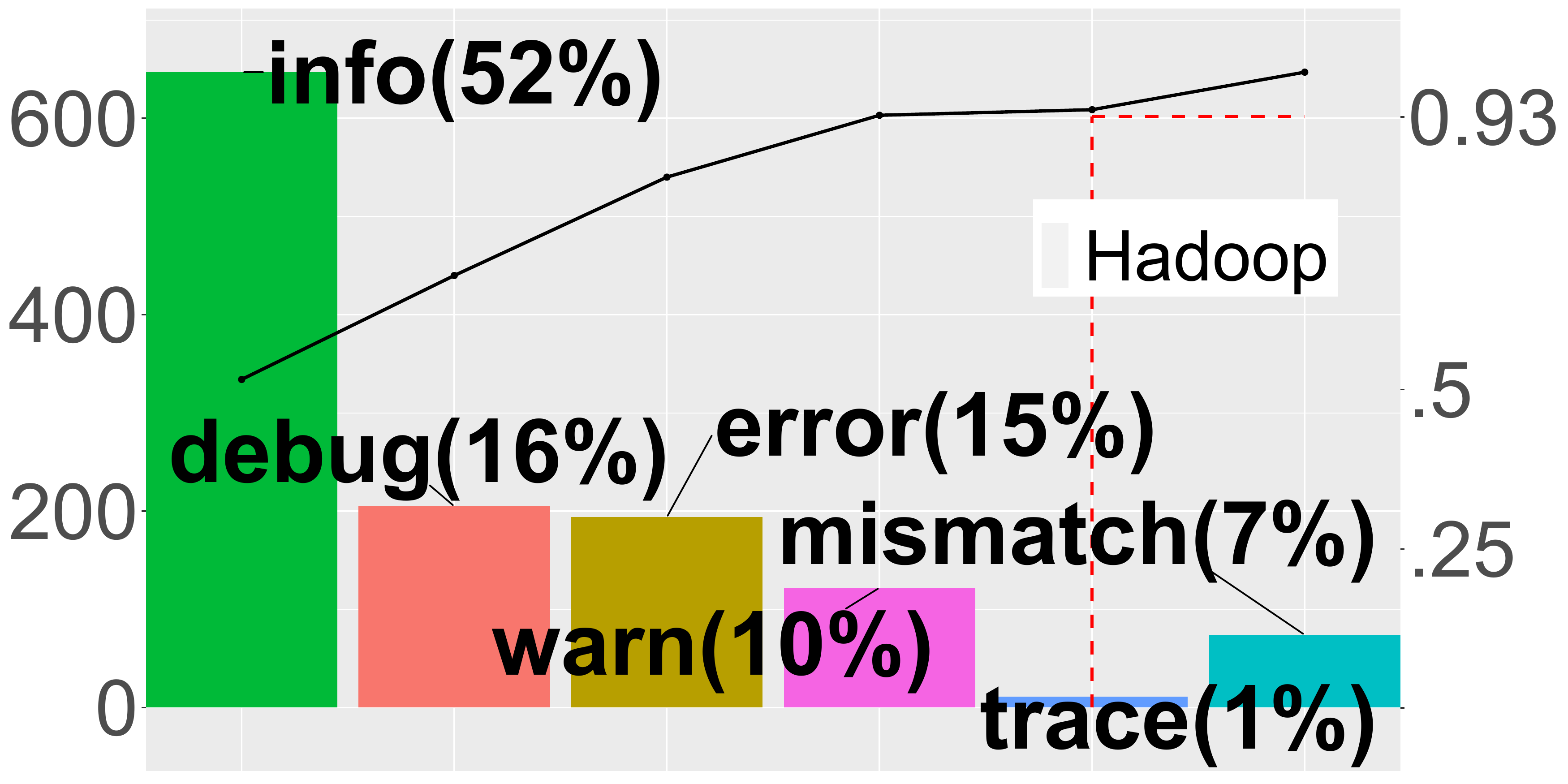}
        \end{subfigure} 
        \begin{subfigure}[b]{0.23\textwidth}
            \includegraphics[width=1\textwidth]{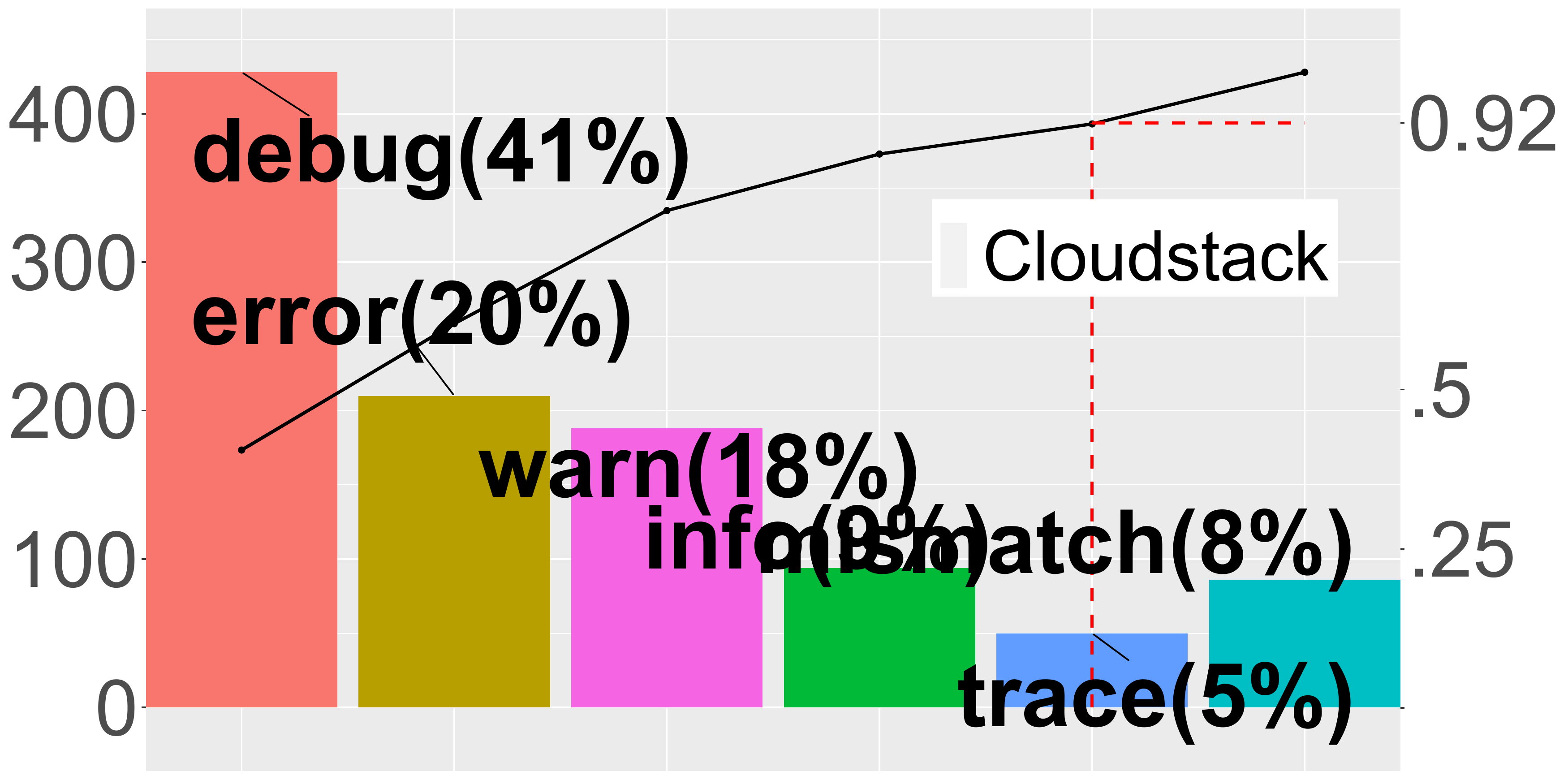}
        \end{subfigure}  
               \begin{subfigure}[b]{0.23\textwidth}
            \includegraphics[width=1\textwidth]{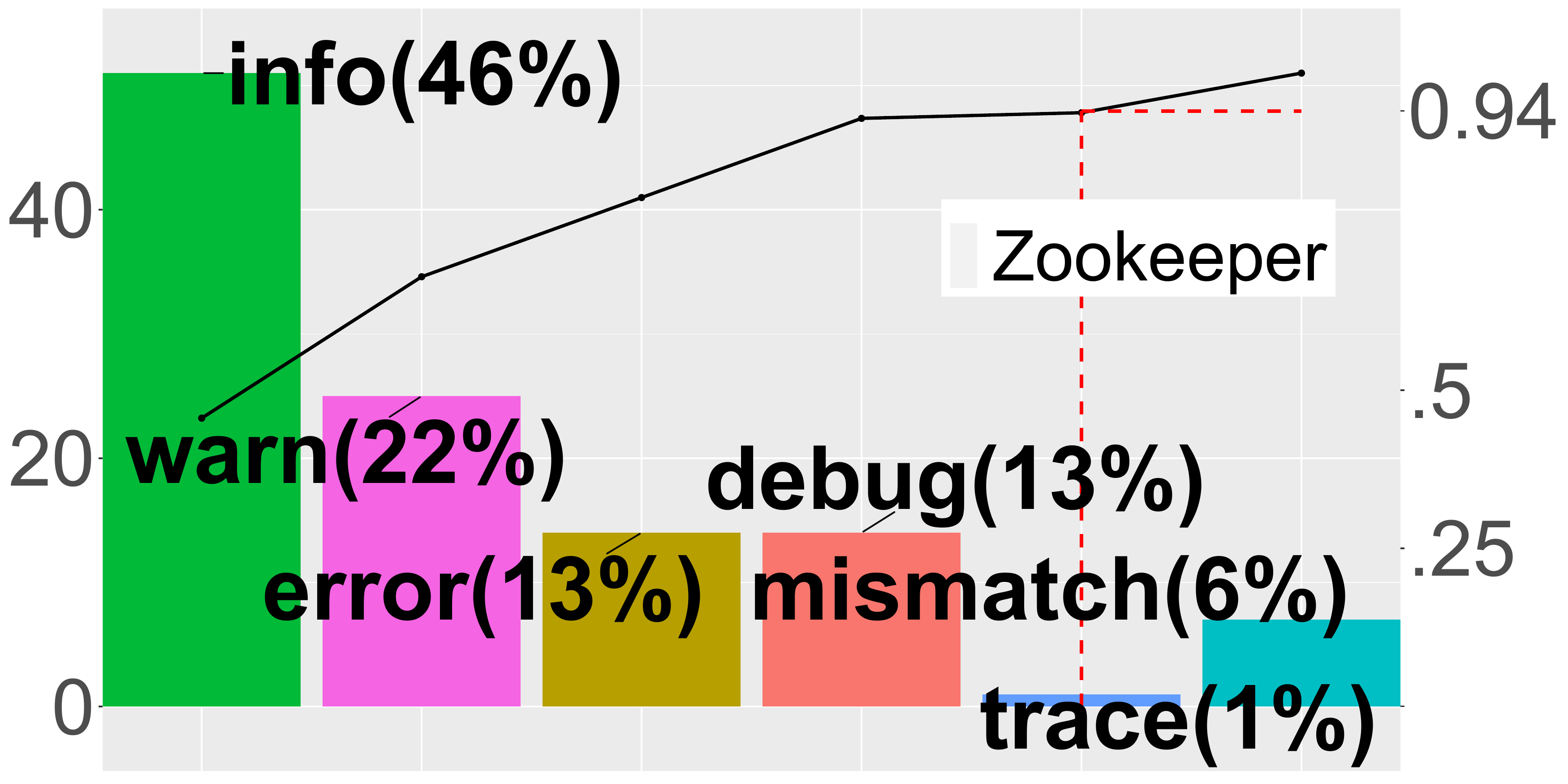}
        \end{subfigure}  
         \caption[ ]{Percentage values for each verbosity level. For each project, only a small percentage of clones have a \textit{`mismatch'} in their log verbosity levels.} 
        \label{fig_lvl}
\end{figure}
}
\subsection{Practicality in Software Engineering}
We note that the ideas similar to our approach for automated log generation have been already applied and proven to be effective in adjacent software engineering tasks such as automated \textit{commit message}~\cite{wang2021context} and \textit{comment}~\cite{wei2020retrieve} generation. 
For example, Wei \textit{et al.}~\cite{wei2020retrieve} used comments of similar code snippets as \textit{`exemplars'} to assist in generating comments for new code snippets. 
Both papers' ideas and application scenarios are analogous to a large extent to those of our work. 
Similarly, both approaches utilize BLEU~\cite{wei2020retrieve,wei2020retrieve} and ROUGE-L~\cite{wei2020retrieve} scores for evaluating the quality of the auto-generated text.

\section{Threats to validity}\label{threats}
We categorize external and internal threats to the validity of our research. 
\subsection{External Threats}
External threats to the validity reflect on the generalization of our work to other such software projects and programming languages. 
In this research, we conducted our log statement analysis on seven open-source Java projects that are well-established and continuously maintained, and used in prior logging research~\cite{chen2017characterizing,he2018characterizing}. 
We assumed our approach is independent of the underlying programming language that the source code is implemented with.    
However, since other software systems, and other programming languages, may follow different logging practices, our findings may not accurately extend and generalize to other such systems.
\subsection{Internal Threats}
Regarding internal threats, our approach relies on the clone detector to find a clone pair for providing logging suggestions, which implies that we cannot suggest a logging statement for a newly-developed code snippet if it is not similar to a priorly-developed code snippet. 
In fact, we argue this threat is not exclusive to our approach but also exists for all other log prediction approaches that rely on learning from logged snippets of source code~\cite{zhu2015learning,li2020shall}, as a learning model can only predict a logging statement for a new code snippet if it can find a feature mapping to its learned logged code base. 
%, assuming this new code snippet requires a logging statement.
To mitigate this concern, we suggest curating a database of available open-source code that can be readily parsed and become available for clone detection. 
This database could be used to improve the \textit{hit-rate} of the clone detection approach when searching for similar code snippets~\cite{hindle2012naturalness,he2018characterizing}. 
Additionally, the architecture of the LSTM model and tuning of its hyperparameters~\cite{greff2016lstm} can have an impact on the BLEU and ROUGE scores for different software projects.

\section{Related Work}\label{rwork}
We categorize the prior work into three main areas: log prediction, code clone detection, and NLP research in software systems.

\subsection{Log Prediction}
Yuan \textit{et al.} proposed \textit{ErrorLog}~\cite{yuan2012conservative}, a tool to report error handling code, \textit{i.e.}, \textit{error logging}, such as \textit{catch clauses}, which are not logged, and to improve the code quality and help with failure diagnosis by adding log statements to these unlogged code snippets. 
Zhao \textit{et al.}~\cite{zhao2017log20} introduced \textit{Log20}, a performance-aware tool to inject new logging statements to the source code to disambiguate execution paths. 
\textit{Log20} introduces a logging mechanism that does not consider developers' logging habits or concerns. 
Zhu \textit{et al.}~\cite{zhu2015learning} proposed \textit{LogAdvisor}, a learning-based framework, for automated logging prediction which aims to learn the frequently occurring logging practices automatically. 
Their method learns logging practices from existing code repositories for \textit{exception} and \textit{function return-value check} blocks by looking for textual and structural features within these code blocks with logging statements.    
Jia \textit{et al.}~\cite{jia2018smartlog} proposed an intention-aware log automation tool called \textit{SmartLog}, which uses an \textit{Intention Description Model} to explore the intention of existing logs. 
Li \textit{et al.}~\cite{li2020shall} categorized six block-level logging locations, and C\^andido \textit{et al.}~\cite{candido2021exploratory} performed an exploratory study of log placement with transfer learning for an enterprise software. 
We discussed how our approach differs from these works during the discussion in Section~\ref{logprediction}. 
Also, our research is the only one that tackles both log location and description automation. 
Gholamian and Ward~\cite{gholamian2020logging} showed that code clones follow similar logging patterns and investigated the feasibility of predicting the \textit{``location''} of log statements in an experimental study, but failed to fully observe the clone-detection shortcomings for log prediction.
Another research also proposed steps involved in leveraging similar code snippets for log statement prediction~\cite{gholamian2021leveraginganonymous}. 
The author also discusses the practicality of their approach during the software's development cycle.
Our goal in this paper is to improve on the performance of log-aware clone detection and also predict the \textit{``description''} of log statements by utilizing code clones and deep-learning NLP approaches.

\subsection{Code Clone Detection}
Source code clone detection is a well-established area of study for software systems, and a significant number of detection techniques and tools have been presented in the literature~\cite{rattan2013software,ain2019systematic}. 
Code-clone detection is the task of identifying syntactically exact or similar snippets of source code (with equal semantics) within or between software systems~\cite{saini2018oreo,sajnani2016sourcerercc} based on contextual features of the source code snippets. 
We demonstrated that searching for similar code snippets, \textit{i.e.}, clone pairs, is beneficial in automated log statement generation. 
We initially observed the shortcomings of the generic state-of-the-art clone detection methods~\cite{saini2018oreo} for log automation, and then improved on by our log-aware method (LACCP), and proposed an approach to suggest log statements' \textit{description} (NLP CC'd).

\subsection{NLP in Software Systems}
%\textbf{NLP in software systems.} 
Prior research has widely utilized natural language attributes for various applications in software engineering. 
For example, natural language exists in software source code and identifier names, design documents, bug reports~\cite{arnaoudova2015use}, and code suggestion~\cite{bhoopchand2016learning}. 
To enable NLP for software systems, prior research \cite{gabel2010study,hindle2012naturalness} has shown the source code is redundant and repetitive, which can be utilized to model the source code with n-gram language models. 
Tu \textit{et al.} ~\cite{tu2014localness} further explored the localness of software characteristics in order to utilize regularities that can be captured in a locally estimated cache and leveraged for software engineering tasks. 
Most recently, research has shown that logging statements' descriptions~\cite{he2018characterizing} and execution log files~\cite{gholamian2021naturalness} manifest natural language features, similar to other software artifacts, such as source code itself. 
Inspired by the prior NLP research, in this work, we utilize a deep-learning NLP model for logging statements' description prediction, and outperform the results in prior work~\cite{he2018characterizing}.

\section{Conclusions and Future Directions}\label{conc}
Software developers insert logging statements in the source code in various places to improve the software development process and its diagnosability. 
Nevertheless, this process is currently manual, and it does lack a unified guideline for the location and content of log statements. 
In this paper, with the goal of log automation, we presented a study on the \textit{location} and \textit{description} of logging statements in open-source Java projects by applying similar code snippets and NLP models.
We initially improved the performance of the log-aware code clone detector (LACCP) by 15.6\% compared to Oreo, and then augmented the performance of log description prediction with the deep learning natural language processing approaches. 
We experimented on seven open-source Java systems, and our analysis shows that by utilizing log-aware clone detection and NLP, our hybrid model, (\textit{NLP CC'd}), achieves 40.86\% higher performance on BLEU and ROUGE scores for predicting LSDs when compared to the prior research (Z{\tiny(over)}W), and achieves 6.41\% improvement over the No-NLP version (Z{\tiny(over)}X). 
We also included a case study of logging issues in Hadoop, a discussion on the applicability of our approach and prediction of the log verbosity level and its variables, and threats to the validity of our research.
As future work, we look into further incorporating the source code surrounding the logging statements for additional context in log automation.

\bibliographystyle{ACM-Reference-Format}
\bibliography{sigproc}

\appendix

\section{Repository Explained}
We provide a repository~\cite{rep_package_laccp} to make our data available. 
The main folders are \textit{LACCPlus} and \textit{NLPCCd} for RQ1 and RQ2, respectively. 
Under each folder, there are subfolders for each software project, \textit{e.g.}, \textit{Zookeeper}. 
Inside each subfolder, we have provided clone pairs for methods that we have examined in our study.  
The naming convention for each method consists of its id (\textit{i.e.}, method id) and an index for each method snippet, such that \textit{(id\_1, id\_2)} forms a clone pair.

\end{document}